\documentclass[prd,nofootinbib,preprint,superscriptaddress,secnumarabic]{revtex4}
\pdfoutput=1

\ifx\pdfoutput\undefined
\usepackage[dvips,bookmarks=false]{hyperref}	
\else
\usepackage{hyperref}	
\fi
\hypersetup{colorlinks,bookmarksopen,bookmarksnumbered,citecolor=blue,
linkcolor=black,pdfstartview=FitH,urlcolor=blue}

\usepackage{graphicx}
\usepackage{amsmath}
\usepackage{amssymb}

\newcommand{\be}{\begin{equation}}
\newcommand{\ee}{\end{equation}}
\newcommand{\beq}{\begin{equation}}
\newcommand{\eeq}{\end{equation}}

\newcommand{\vect}[1]{\boldsymbol{\rm #1}}
\newcommand{\ev}[1]{\ensuremath{\left\langle #1 \right\rangle}} 

\newcommand{\vhalo}{\vect{\hat v}_{\rm halo}}

\newcommand{\capdef}{}
\newcommand{\mycaption}[2][\capdef]{\renewcommand{\capdef}{#2}%
       \caption[#1]{{\footnotesize #2}}}
\makeatletter
\renewcommand{\fnum@table}{\textbf{\tablename~\thetable}}
\renewcommand{\fnum@figure}{\textbf{\figurename~\thefigure}}
\makeatother

\begin{document}
\pagestyle{plain}

\vspace*{1cm}

\title{
On the annual modulation signal in dark matter direct detection
\vspace*{1cm}
}

\def\ific{Departamento de Fisica Teorica, and IFIC, Universidad de Valencia-CSIC
Edificio de Institutos de Paterna, Apt. 22085, 46071 Valencia, Spain}
\def\mpi{Max-Planck-Institut f{\"u}r Kernphysik, PO Box 103980, 69029 Heidelberg, Germany}
\author{\textbf{Juan Herrero-Garcia}\vspace*{0mm}}
\email{juan.a.herrero AT uv.es}
\affiliation{\mpi}
\affiliation{\ific}

\author{\textbf{Thomas Schwetz}\vspace*{0mm}}
\email{schwetz AT mpi-hd.mpg.de}
\affiliation{\mpi}

\def\Cincy{Department of Physics, University of Cincinnati, Cincinnati, Ohio 45221,USA}
\author{\textbf{Jure Zupan\footnote{On leave of absence from University of Ljubljana, Depart. of Mathematics and Physics, Jadranska 19, \\[-1mm]
 1000 Ljubljana, Slovenia and Josef Stefan Institute, Jamova 39, 1000 Ljubljana, Slovenia.}}\vspace*{3mm}}
\email{jure.zupan AT cern.ch}
\affiliation{\Cincy}


\begin{abstract}
\vspace*{5mm} 

We derive constraints on the annual modulation signal in Dark Matter (DM)
direct detection experiments in terms of the unmodulated event rate. A
general bound independent of the details of the DM distribution follows from the
assumption that the motion of the earth around the sun is the only source of
time variation. The bound is valid for a very general class of particle physics
models and also holds in the presence of an unknown unmodulated background.
More stringent bounds are obtained, if modest assumptions on symmetry
properties of the DM halo are adopted. We illustrate the bounds by applying
them to the annual modulation signals reported by the DAMA and CoGeNT
experiments in the framework of spin-independent elastic scattering. While
the DAMA signal satisfies our bounds, severe restrictions on the DM mass can
be set for CoGeNT.
\end{abstract}
\maketitle

\section{Introduction}
\label{sec:intro}

A smoking-gun signature of the dark matter (DM) signal in DM direct
detection experiments is the annual modulation of the event rate. This
arises because the earth rotates around the sun, while at the same
time the sun moves relative to the DM halo~\cite{Drukier:1986tm,
  Freese:1987wu}. Currently, two experiments report annual modulation
of the signals, DAMA with 8.9$\sigma$
significance~\cite{Bernabei:2008yi}, and CoGeNT with a significance of
$2.8\sigma$~\cite{Aalseth:2011wp}.  An important question is whether
the observed modulation rates are consistent with the interpretation
that the signal is due to DM.  Ideally, one would like to answer this
question without requiring a detailed knowledge of the DM halo. As we
will show below, we come very close to this ideal.
 
The DM scattering rate in the detector is determined by the particle physics
properties of DM and by the astrophysical properties of the DM halo. The
latter carry large uncertainties. When presenting results of DM direct
detection experiments, it has become customary to use a ``standard"
Maxwellian DM velocity distribution in order to show the constraints on DM
mass and scattering cross section. This is very likely an
oversimplification, with N-body simulations indicating a more complicated
structure of the DM halo, see e.g., \cite{Kuhlen:2009vh}.  It is well known
that the annual modulation signal is sensitive to such deviations from the
Maxwellian halo \cite{Fornengo:2003fm, Green:2003yh,Fairbairn:2008gz}. DM
streams are particularly relevant for the annual modulation signal and can
lead to large effects. These can, for instance, be comparable in size to the
modulation effect due to the whole Maxwellian halo \cite{Gelmini:2000dm,
Savage:2006qr, Chang:2008xa, Natarajan:2011gz}. 

In the following we will derive constraints on the amplitude of the annual
modulation using measured unmodulated rates.  The derivation of these bounds
is made possible by the fact that the velocity of the earth rotating around the
sun, $v_e\sim 30$ km/s, is much smaller than the typical velocity of DM
particles in the halo, $\langle v \rangle \sim 200$~km/s, and we can expand
the velocity integral to first order in $v_e$. The derived constraints
on the annual modulation then serve as a consistency check for the hypothesis
that the annual modulation is due to the DM signal. The check that we
provide is (almost) independent of DM halo properties.  Previously halo
independent methods to interpret DM direct detection data have been
developed and applied in \cite{Drees:2007hr, Drees:2008bv, Fox:2010bz,
Fox:2010bu, McCabe:2011sr, Frandsen:2011gi}, with different goals than ours.
In ref.~\cite{Drees:2008bv} a method for determining the DM mass
independently of the halo was developed, while refs.~\cite{Drees:2007hr,
Frandsen:2011gi} focused on extracting halo properties from the data.
Compatibility studies of different experiments without adopting assumptions
on the halo have been performed in~\cite{Fox:2010bz, Fox:2010bu,
McCabe:2011sr}. Our goal is to obtain a consistency check between the
modulation amplitude and the unmodulated signal within a given experiment.

The structure of the paper is as follows: in sec.~\ref{sec:notation}
we set up the notation, followed in sec.~\ref{sec:general} by a
derivation of the general bound on the modulation signal under the very
mild assumption that the properties of the DM halo in our vicinity do
not change on time scales of months. In sec.~\ref{sec:spec} we impose
further symmetry requirements on the DM velocity distribution and
obtain successively more stringent bounds. In sec.~\ref{sec:data} we
then demonstrate the usefulness of our bounds by applying them to the
modulation signals reported in CoGeNT and DAMA within elastic
spin-independent scattering. We find that our method leads to
non-trivial restrictions on DM interpretations of the modulation
signal in CoGeNT whereas the DAMA signal satisfies our bounds for the
relevant DM masses. We summarize in sec.~\ref{sec:conclusion}. In
appendix~\ref{sec:maxw} we consider the Maxwellian DM halo and show
that the expansion in $v_e$ is rather accurate in this
case. Furthermore, we illustrate the various bounds compared to the
modulation signal expected for the Maxwellian
halo. Appendix~\ref{sec:translation} provides technical details
necessary to translate the bounds derived for the halo velocity
integral into observable event rates. Supplementary material related
to the treatment of an unknown background is given in
appendix~\ref{app:proc2}.

\section{Notation}
\label{sec:notation}

The differential rate in events/keV/kg/day for DM $\chi$ to scatter
off a nucleus $(A,Z)$ and depositing the nuclear recoil energy
$E_{nr}$ in the detector is
\beq \label{rate}
{R}(E_{nr},t) = \frac{\rho_\chi}{m_\chi} \frac{1}{m_A}\int_{v>v_{m}}d^3 v \frac{d\sigma_A}{d{E_{nr}}} v f_{\rm det}(\vect v, t).
\eeq
Here $\rho_\chi \simeq 0.3 \, {\rm GeV/cm}^3$ is the local DM density, $m_A$
and $m_\chi$ are the nucleus and DM masses, $\sigma_A$ the DM--nucleus
scattering cross section\footnote{Throughout this work we
assume that DM is dominated by a single particle species. A generalization
to multi-component DM is straightforward and will be pursued in future work.} 
and $\vect v$ the 3-vector relative velocity
between DM and the nucleus, while $v\equiv |\vect{v}|$. For a DM particle to
deposit recoil energy $E_{nr}$ in the detector a minimal velocity $v_{m}$ is
required, restricting the integral over velocities in \eqref{rate}. For
elastic scattering it is

\beq\label{eq:vm}
v_{m}=\sqrt{ \frac{m_A E_{nr}}{2 \mu_{\chi A}^2}}\,,
\eeq
where $\mu_{\chi A}$ is the reduced mass of the DM--nucleus system. 

The function $f_{\rm det}(\vect v, t)$ describes the distribution of DM particle velocities in the detector rest frame with $f_{\rm det}(\vect v, t) \ge 0$ and
$\int d^3 v f_{\rm det}(\vect v, t) = 1$. It is related to the velocity distribution 
in the rest frame of the sun by 
\beq 
f_{\rm det}(\vect{v},t) = f_{\rm sun}(\vect{v} + \vect{v}_e(t)) \,,
\eeq 
where $\vect{v}_e(t)$ is the velocity vector of the earth, which
we write as \cite{Gelmini:2000dm}
\beq\label{eq:ve}
\vect{v}_e(t) = v_e [\vect{e}_1 \sin\lambda(t) - \vect{e}_2
\cos\lambda(t) ]
\eeq
with $v_e=29.8$ km/s, and $\lambda(t)=2\pi(t-0.218)$ with $t$ in
units of 1 year and $t=0$ at January 1st, while $\vect e_1 = (-0.0670,0.4927,-0.8676)$ and
$\vect e_2 =(-0.9931,-0.1170,0.01032)$ are orthogonal unit
vectors 
spanning the plane of the earth's orbit, assumed to be circular.
Similarly, the DM velocity distribution in the galactic frame is
connected to the one in the rest frame of the sun by $ f_{\rm
  sun}(\vect{v}) = f_{\rm gal}(\vect{v} + \vect{v}_{\rm sun})$, with
$\vect v_{\rm sun} \approx (0,220,0) \, {\rm km/s} + \vect v_{\rm
  pec}$ and $\vect v_{\rm pec} \approx (10,13,7)$~km/s the peculiar
velocity of the sun. We are using galactic coordinates where $x$
points towards the galactic center, $y$ in the direction of the galactic
rotation, and $z$ towards the galactic north, perpendicular to the
disc. As shown in \cite{Green:2003yh}, eq.~\eqref{eq:ve} provides an
excellent approximation to describe the annual modulation signal.

In the following we consider the typical situation, 
where the differential cross section is given by 
\begin{align}
  \frac{d\sigma_A}{dE_{nr}} = \frac{m_A}{2\mu_{\chi A}^2 v^2} \sigma_A^0 F^2(E_{nr}) \,, 
  \label{eq:dsigmadE}
\end{align}
where $\sigma_A^0$ is the total DM--nucleus scattering cross section at zero momentum transfer, and $F(E_{nr})$ is a form factor. The event rate is then given by
\beq\label{eq:Rgamma}
R(E_{nr}, t) = C \, F^2(E_{nr}) \, \eta(v_m, t) 
\qquad\text{with}\qquad
C = \frac{\rho_\chi \sigma_A^0}{2 m_\chi \mu_{\chi A}^2}
\eeq
and the halo integral
\beq\label{eq:eta} 
\eta(v_m, t) \equiv \int_{v > v_m} d^3 v \frac{f_{\rm det}(\vect{v}, t)}{v} \,.
\eeq
Here and in the following $v_m$ and $E_{nr}$ are related by eq.~\eqref{eq:vm}.
This formalism covers a wide range of possible DM--nucleus interaction models, including the standard spin-independent and spin-dependent scattering. The results we derive below 
apply to all the cases where $d\sigma_A/dE_{nr} \propto 1/v^2$, in which case the halo integral \eqref{eq:eta} is obtained. The arguments can easily be generalized to a non-trivial $q^2$ dependence of the interaction, which would introduce an additional $E_{nr}$ dependent function in \eqref{eq:Rgamma} but would not change \eqref{eq:eta}. Our results do not apply, however, to a non-standard velocity dependence of the cross section, which would modify eq.~\eqref{eq:eta}.

The rate will have a time independent component and an annually modulated component which we define as 
\beq
R(E_{nr},t)= \overline R(E_{nr})+\delta R(E_{nr},t) = 
C \, F^2(E_{nr}) \, [ \overline \eta (v_m) + \delta\eta(v_m, t)] \,.
\eeq
Below we will be specifically interested in purely sinusoidal time
dependence with period of one year, in which case we can write
\beq\label{eq:modulation}
\begin{array}{c}
\delta R(E_{nr}, t) = A_R(E_{nr}) \cos 2\pi[t - t_0(E_{nr})] \,, \\
\delta \eta(v_m, t) = A_\eta(v_m) \cos 2\pi[t - t_0(E_{nr})] \,.
\end{array}
\eeq
The peak of the annual modulation occurs at $t_0$, which in general depends 
on $E_{nr}$ (or equivalently on $v_m$). The modulation amplitudes of the event rate, $A_R$,
and of the halo integral, $A_\eta$, are related through $A_R(E_{nr}) = C
F^2(E_{nr}) A_\eta(v_m) \ge 0$. We will first derive bounds on $A_\eta$ in
terms of the time averaged value of the halo integral $\overline\eta$. In
section~\ref{sec:data} we will then translate the bounds into constraints
involving the observable quantities $A_R$ and $\overline R$.

\section{General bound on the annual modulation amplitude}
\label{sec:general}

{\bf Assumption 1:} We assume that the only time dependence comes from
$\vect{v}_e(t)$ and there is no explicit time dependence in $f_{\rm sun}$.

\bigskip

This assumption implies that the halo is spatially constant at the scale of
the sun--earth distance and also constant in time on scale of months. Under this assumption we can derive a general bound on the annual modulation by expanding eq.~\eqref{eq:eta} in $v_e/v_m
\ll 1$ \footnote{Typically $v_m$ is sensitive to both the nucleus and the DM mass. For a $10$ GeV DM mass and a recoil energy $E_{nr}$ of a few KeV, we get for the nuclei we have analysed (Ge, Na, I) that $v_m \gtrsim 5 v_e$ km/s, so the expansion is rather accurate (as we have checked explicitly for a Maxwellian halo in appendix~\ref{sec:maxw}).}.  Using for short $f \equiv f_{\rm sun}$, we have 
\begin{align}
\eta(v_m&,t) = \int_{|\vect{v} - \vect{v}_e| > v_m} d^3 v 
        \frac{f(\vect{v})}{|\vect{v} - \vect{v}_e|} \label{eq:R2}\\
     &= \int_{v > v_m} d^3 v \frac{f(\vect{v})}{v} 
      + \int d^3 v  f(\vect{v})
             \frac{\vect{v}\cdot\vect{v}_e(t)}{v^3}
            \left[\Theta(v-v_m) - \delta(v - v_m) v_m \right]  + 
            \mathcal{O}(v_e^2/v_m^2) \,, \label{eq:R3}
\end{align}
where the first term in \eqref{eq:R3} gives the time independent part of the
DM scattering signal. In polar coordinates the time independent halo
integral is then given by
\beq \label{eq:unmod}
\overline{\eta}(v_m) =\int_{v > v_m} d^3 v \frac{f(\vect{v})}{v}  
= 
\int_{v_m} dv \, v \int_0^{2\pi} d\varphi \int_{-1}^1
       d\cos\vartheta  \, f(v,\vartheta,\varphi) \,.
\eeq
The second term in \eqref{eq:R3} corresponds to the time dependent part of the DM
scattering signal, with $\vect{v}_e(t)$ given in \eqref{eq:ve}. Expanding
to linear order in $v_e$ thus leads to an annual modulation signal that has
a sinusoidal shape. One can check experimentally for the convergence of the
expansion by searching for higher order terms, $\propto v_e^2
\sin^2[2\pi(t-t_0)]$, etc. Note that the expansion is in $v_e/v$, where $v
\gtrsim v_m$, so that the accuracy is typically better than ${\mathcal
O}(v_e/v_m)$. In appendix~\ref{sec:maxw} we demonstrate explicitly that this
expansion is rather accurate in the case of a Maxwellian halo. 

By expanding the velocity integral in the small quantity
$v_e$ we assume that $f(\vect{v})$ is smooth on scales $\lesssim v_e$.
Hence, our bounds do not apply in situations where $f(\vect{v})$ has strong
structures at scales smaller than $v_e$. An example would be a very cold
stream with velocity $v_\mathrm{stream}$ and a dispersion smaller than
$v_e$. In such a case the expansion will not be accurate for $v_m$ in the
range $|v_m - v_\mathrm{stream}| \lesssim v_e \sin\alpha$, though it may
still work to good accuracy for $v_m$ outside this range.

The time dependent component in \eqref{eq:R3}
has two contributions 
\beq
\delta\eta(v_m,t)=\int d^3 v  f(\vect{v})
             \frac{\vect{v}\cdot\vect{v}_e(t)}{v^3}
            \left[\Theta(v-v_m)- \delta(v - v_m) v_m \right].
\eeq
The term with the $\Theta$-function comes from expanding the denominator in
eq.~\eqref{eq:R2} and involves a volume integral over the region $v > v_m$.
The term with the $\delta$-function comes from taking into account
the effect of $v_e$ on the integration boundary, and the $\delta$-function
indicates that the argument has to be evaluated at the surface $v = v_m$. Let us treat the
two terms separately and define 
\begin{align}
\vect{\hat v}_g(v_m) g(v_m)\equiv \int d^3 v & f(\vect{v}) \frac{\vect{v}}{v^3} \delta(v - v_m)  
\,, \label{eq:g1}\\
 \vect{\hat v}_G(v_m) G(v_m)\equiv \int d^3 v  & f(\vect{v}) \frac{\vect{v}}{v^3} \Theta(v - v_m)  
 \,. \label{eq:G1}
\end{align}
The unit vectors $\vect{\hat v}_g(v_m)$ and $\vect{\hat v}_G(v_m)$ give the
corresponding weight averaged DM wind directions in the earth's rest frame. In
general they point in different directions and can depend on $v_m$. For
the Maxwellian halo they point in the same direction and are equal to
$\vect{\hat v}_g=\vect{\hat v}_G=-\vect{v}_{\rm sun}$. We will treat such
special cases in the next section. The positive functions $g(v_m)$ and
$G(v_m)$ are given by
\begin{align}
g(v_m) &= \int_0^{2\pi} d\varphi \int_{-1}^1
       d\cos\vartheta f(v_m,\vartheta,\varphi) \cos\vartheta \,, \label{eq:g}\\
       G(v_m) &= \int_{v_m} dv \int_0^{2\pi} d\varphi \int_{-1}^1
       d\cos\vartheta' f(v,\vartheta',\varphi) \cos\vartheta' \,, \label{eq:G}
\end{align}
where $\vartheta (\vartheta')$ is the
angle between $\vect{v}$ and $\vect{\hat v}_g (\vect{\hat v}_G)$.
The time dependent halo integral is thus given by 
\beq
\delta \eta(v_m, t) = \vect{v}_e(t) \cdot \left[
\vect{\hat v}_G(v_m) G(v_m)-\vect{\hat v}_g(v_m) v_m g(v_m)\right] \,. \label{eq:M}
\eeq
As already mentioned, the form of $\vect{v}_e(t)$ from eq.~\eqref{eq:ve} implies that our
approximations lead to strictly sinusoidal modulations, justifying the
ansatz in eq.~\eqref{eq:modulation}, $\delta \eta(v_m, t) = A_\eta(v_m) \cos 2\pi(t - t_0)$. 
Using eq.~\eqref{eq:M}, the modulation amplitude can be constrained in the
following way:
\beq \label{eq:M1}
A_\eta(v_m) \leq v_e \big[v_m g(v_m)+G(v_m)\big] \,.
\eeq
Note that the two terms in eq.~\eqref{eq:M} will contribute to the
modulation amplitude proportional to $\sin\alpha_{g}$ and $\sin\alpha_{G}$, 
respectively, where $\alpha_{g,G}$ is the angle between $\vect{\hat
v}_{g,G}$ and the direction orthogonal to the plane of the earth orbit,
i.e., $\cos\alpha_{g,G} = \vect{\hat v}_{g,G} \cdot \vect{e_3}$, with
$\vect{e_3} = \vect{e_1} \times \vect{e_2}$, and $\vect{e}_{1,2}$ given
below eq.~\eqref{eq:ve}. In general $\alpha_{g,G}$ will depend on $v_m$. To
derive the bound we have assumed the maximal possible effect, corresponding
to $\sin\alpha_g = \sin\alpha_G = 1$. Since we do not know the relative sign
of the contributions from $g(v_m)$ and $G(v_m)$ we have to take the sum of
the moduli. 

The function $g(v_m)$ is bounded from above by 
\beq \label{eq:bound_g}
g(v_m) \leq \int d\varphi \,d\negthinspace \cos\vartheta f(v_m, \varphi, \vartheta) 
   |\cos\vartheta|
 \leq \int d\varphi \,d\negthinspace\cos\vartheta f(v_m, \varphi, \vartheta) 
  = - \frac{1}{v_m}\frac{d \overline{\eta}}{d v_m} \,,
  \eeq
where the last equality follows from eq.~\eqref{eq:unmod}. Note that
$\overline{\eta}(v_m)$ is a monotonously decreasing function and therefore,
$d\overline{\eta}(v_m) / dv_m \leq 0$.  Using again the last equality above, 
also a bound for $G(v_m)$ can be derived:
\begin{align} 
G(v_m) \leq 
   - \int_{v_m} dv \frac{1}{v}\frac{d \overline{\eta}}{d v}  
   &= \frac{\overline{\eta}(v_m)}{v_m}  
   - \int_{v_m} dv \frac{\overline{\eta}(v)}{v^2}. \label{eq:bound_G_impr}
\end{align}
The equality in \eqref{eq:bound_G_impr} follows from integration by parts.
The inequalities \eqref{eq:bound_g} and \eqref{eq:bound_G_impr} are
saturated if $f(\vect{v}) \propto \delta(\vartheta)$, i.e., the hypothetical
situation that all DM particles have the same direction and their velocities
have no transversal component. Using eqs.~\eqref{eq:bound_g} and
\eqref{eq:bound_G_impr} the bound on the modulation amplitude \eqref{eq:M1}
becomes
\beq \label{eq:bound_gen}
A_\eta(v_m) \leq v_e 
\left[-\frac{d \overline \eta}{d v_m}+ 
    \frac{\overline{\eta}(v_m)}{v_m}  
   - \int_{v_m} dv \frac{\overline{\eta}(v)}{v^2} 
\right] 
\qquad\text{(Assumption 1)}\,,
\eeq
where the first two terms on the r.h.s.\ are positive and the third is
negative. If the DM scattering rate $\overline R$ is measured, the r.h.s.\
is fully determined experimentally by $\overline \eta(v_m) = \overline
R(E_{nr}) / C F^2(E_{nr})$, and can be compared to the observed modulation
through $A_\eta = A_R / C F^2(E_{nr})$, see section~\ref{sec:data} for
details. Note that the phase of the modulation (which may vary with $v_m$)
does not appear in the bound \eqref{eq:bound_gen}. The bound applies on the
modulation amplitude, irrespective of the phase. Eq.~\eqref{eq:bound_gen} is
one of the main results of this paper. The bound is rather general and holds
under the very mild Assumption~1 specified above. For it to be saturated the
DM velocity distribution at the position of the solar system would need to
be highly nontrivial. For instance, the bound in \eqref{eq:M1} can be
saturated if there is a DM stream aligned with the ecliptic and it is strong
enough so that it dominates the velocity distribution at $v=v_m$. Note that
even in this case, for different $v_m$ the bound will not be saturated. To
saturate in addition eq.~\eqref{eq:bound_g} or \eqref{eq:bound_G_impr}, the
stream should have no transversal velocity dispersion. Any modulation signal
which violates, or even just saturates, this bound is very unlikely to have
a DM origin.

\section{Bounds on the annual modulation for symmetric halos}
\label{sec:spec}

{\bf Assumption 2:} In addition to Assumption 1 we now assume that
$\vect{\hat v}_g$ defined in eq.~\eqref{eq:g1} is independent of $v_m$,
i.e., we assume that there is a constant direction $\vhalo \equiv \vect{\hat v}_g$
governing the shape of the DM velocity distribution in the sun's rest
frame.

\bigskip

From this assumption it follows immediately that $\vect{\hat v}_G$ is also constant and equal to $\vhalo$, so that
\beq \label{eq:deriv}
g(v_m) = - \frac{d G}{d v_m} \,,
\eeq
and eq.~\eqref{eq:M} becomes 
\beq
\delta \eta(v_m, t) = - \vect{v}_e(t) \cdot \vhalo 
\left[v_m g(v_m)-G(v_m)\right] \,. \label{eq:M5}
\eeq
The crucial difference to the general case \eqref{eq:M} is that we were able
to pull the velocity vector in front of the bracket. The functions $v_m
g(v_m)$ and $G(v_m)$ are both positive. Their relative sizes determine
whether the bracket is positive or negative. For small $v_m$ the function
$G(v_m)$ typically dominates and we get an extra half a year shift in the
peak of the modulation (for the Maxwellian halo the peak would then be in
December). For larger $v_m$ the boundary term $v_m g(v_m)$ dominates and the
whole bracket is positive (and thus for the Maxwellian halo the peak in this
case is in June, see appendix~\ref{sec:maxw}). 

Assumption 2 is fulfilled if $f(\vect{v})$ obeys certain symmetry requirements that we can deduce from eq.~\eqref{eq:g1}. For a given $v_m$ we chose a coordinate system such that $\vect{\hat v}_g = (1,0,0)$, and
\beq\label{eq:cond2}
 \int d^3 v  f(\vect{v}) v_y \delta(v-v_m) = 0 \,,\qquad
 \int d^3 v  f(\vect{v}) v_z \delta(v-v_m) = 0 \,.
\eeq
Assumption 2 implies that these relations hold for any $v_m$ in the same
coordinate system. The distribution in $v_x$ (i.e., in the direction of
$\vhalo$) can be arbitrary, and can for instance even include several peaks, as long as
$\vect{\hat v}_g$ does not flip sign. Therefore, we have to require that the
integral over the half-sphere with $v_x > 0$ is larger than the one with
$v_x < 0$ for all $v_m$: 
\beq \label{eq:cond3}
\int_{v_x < 0} d^3 v f(\vect{v}) v_x \delta(v-v_m) <
\int_{v_x > 0} d^3 v f(\vect{v}) |v_x| \delta(v-v_m) \,. 
\eeq

One possibility to satisfy the condition \eqref{eq:cond2} is a
symmetric velocity distribution, with $f(v_x, v_y, v_z) = f(v_x, -v_y,
v_z)$ and $f(v_x, v_y, v_z) = f(v_x, v_y, -v_z)$ for all
$v_x$. Eq.~\eqref{eq:cond2} can also be satisfied for distributions
asymmetric in $v_y$ and/or $v_z$, however, in this case the asymmetry
has to be such that the cancellation between $v_{y,z} > 0$ and $<0$
happens for all radii $v_m$. 

Assumption~2 is fulfilled for the standard Maxwellian halo, as well as for
any other isotropic velocity distribution. Up to small corrections due to
the peculiar velocity of the sun it holds also for tri-axial halos, and
covers also streams parallel to the motion of the sun, such as a dark disc
co-rotating with the galactic stellar disc \cite{Read:2009iv}. Note that for
all those examples the DM direction $\vhalo$ is aligned with the motion of
the sun (up to the peculiar velocity that leads to sub-leading corrections).
Let us introduce this as an additional assumption:

\bigskip

{\bf Assumption 2a:} In addition to Assumption~2 we require that the
preferred direction $\vhalo$ is aligned with the motion of the
sun relative to the galaxy.

\bigskip

As mentioned above, for many realistic cases fulfilling Assumption~2
also this additional requirement is fulfilled. An exception would be
the situation when the DM density at the sun's location is dominated
by a single stream from an arbitrary direction and the contribution of
the static halo is negligible.

Let us now use eq.~\eqref{eq:M5} to derive a bound on the modulation. We
have $\vect v_e(t)\cdot \vhalo = - v_e \sin\alpha_{\rm halo} \cos(t - t_0)$,
where $t_0$ is now independent of $v_m$.  Here $\alpha_{\rm halo}$ is the
angle between $\vhalo$ and $\vect e_1\times \vect e_2$, i.e., the projection
of $\vhalo$ on the ecliptic plane, with $\sin\alpha_{\rm halo} \ge 0$.
Assumptions~2 and 2a thus imply that the phase of the modulation is
independent of $v_m$ (and therefore independent of $E_{nr}$). As discussed
above this is up to a sign flip of the square bracket in eq.~\eqref{eq:M5}
that can happen due to the two competing terms. To take this into account we
now define
\beq
\delta \eta(v_m, t) = A'_\eta(v_m) \cos(t - t_0) \,, 
\eeq
where $t_0$ is constant and we allow a
sign change for $A'_\eta(v_m)$. This is different from $A_\eta$, which has
been defined to be positive in eq.~\eqref{eq:modulation}. While for
Assumption~2 the phase is arbitrary but constant, Assumption~2a also
gives a prediction for the phase -- that the maximum (or minimum) of the event rate is around
June 2nd. This can be checked in the experiment by looking at the
annual modulation phase at different energy bins. Hence, from the
experimental information on the phase we can conclude whether
Assumptions~2 or 2a may be justified and whether it makes sense to apply the
corresponding test.

A useful bound on the modulation can be obtained by first
integrating eq.~\eqref{eq:M5} over $v_m$, 
\beq
\begin{split}
\int_{v_{m1}}^{v_{m2}} dv_m A'_\eta(v_m)=& v_e \sin\alpha_{\rm halo} 
\int_{v_{m1}}^{v_{m2}} dv_m [v_m g(v_m)- G(v_m)]\\
=& v_e \sin\alpha_{\rm halo} \, \left[v_{m1} G(v_{m1}) - v_{m2} G(v_{m2})\right] \,, \label{eq:intermAeta}
\end{split}
\eeq 
where the second equality is obtained by integrating $v_m g(v_m)$ by
parts using $g(v_m)=-dG/d v_m$, eq.~\eqref{eq:deriv}. Note that the
$A_\eta'$ and $\alpha_{\rm halo}$ are defined in such a way that $A_\eta'$
is positive for large $v_m$, above the last sign flip. Experimentally, this
is at present the most interesting region of the parameter space. Both
putative modulation signals, at CoGeNT and DAMA, have a peak close to June,
and no half year phase change is seen within the observed
energy range.\footnote{Note that the lowest energy bin in
DAMA shows a somewhat smaller modulation amplitude, which might be an
indication of a phase shift below the threshold.}

In general we do not know which of the two $v_mG(v_m)$ terms in
\eqref{eq:intermAeta} dominates. Dropping the smaller of the two and using
the bound \eqref{eq:bound_G_impr} we arrive at 
\begin{align} 
\left|\int_{v_{m1}}^{v_{m2}} \negthickspace \negthickspace \negthickspace dv_m A'_\eta(v_m)\right|
\leq  v_e \big|\negmedspace\sin\alpha_{\rm halo}\big|\,{\rm max}\negthickspace \left[\overline \eta(v_{m1}) - 
v_{m1} \negthickspace \int_{v_{m1}} \negthickspace \negthickspace \negthickspace dv \frac{\overline\eta(v)}{v^2}, v_{m1}\to v_{m2} \right]. \label{eq:bound2}
\end{align}
Assumption~2 allows for arbitrary directions of the DM wind, therefore we
need to use $|\sin\alpha_{\rm halo}|\to 1$ above, while for Assumption 2a
one has $\sin\alpha_{\rm halo}\simeq 0.5$. Note that a sign flip of
$A_\eta'$ will lead to cancellations in the integral on the l.h.s.\ of
eq.~\eqref{eq:bound2} and make the bounds weaker. An observation of such a
sign flip in the modulation amplitude would be a strong experimental
evidence that the modulation is due to a DM signal. If such a sign flip is
observed, one might be able to obtain stronger constraints by applying the
bound for the region below and above the sign flip separately, to avoid the
cancellations.

At present there is no indication of such a sign flip in which case the
bound \eqref{eq:bound2} simplifies to 
\begin{align} 
\int_{v_{m1}}^{v_{m2}} \negthickspace \negthickspace dv_m A'_\eta(v_m)
\leq  v_e\,\left[\overline \eta(v_{m1}) - 
v_{m1} \negthickspace \int_{v_{m1}} \negthickspace \negthickspace
\negthickspace dv \frac{\overline\eta(v)}{v^2} \right] 
\qquad \text{(Assumption 2)}\,, \label{eq:bound_spec}
\end{align}
and
\begin{align} 
\int_{v_{m1}}^{v_{m2}} \negthickspace \negthickspace dv_m A'_\eta(v_m)
\leq  0.5 \, v_e\,\left[\overline \eta(v_{m1}) - 
v_{m1} \negthickspace \int_{v_{m1}} \negthickspace \negthickspace
\negthickspace dv \frac{\overline\eta(v)}{v^2} \right]
\qquad \text{(Assumption 2a)}\,. \label{eq:bound_spec_alpha}
\end{align}

The bounds \eqref{eq:bound_gen}, \eqref{eq:bound_spec},
\eqref{eq:bound_spec_alpha} are the central results of this paper. In
the following we will refer to them as
\begin{center}
\begin{tabular}{l@{\qquad}l@{\qquad}l}
Eq.~\eqref{eq:bound_gen}  & ``general bound'' & (Assumption 1), \\
Eq.~\eqref{eq:bound_spec} & ``symmetric halo'' & (Assumption 2), \\
Eq.~\eqref{eq:bound_spec_alpha} & ``symmetric halo, $\sin\alpha=0.5$'' & (Assumption 2a).
\end{tabular}
\end{center}
The term ``symmetric halo'' should be understood in the sense of
eqs.~\eqref{eq:cond2} and \eqref{eq:cond3}.

For completeness let us also mention an unintegrated bound, even though
we will not use it for the numerical analysis.
Using eq.~\eqref{eq:M5} we have for the modulation amplitude
\beq \label{eq:M1:special} 
A_\eta(v_m) = v_e \big|\sin\alpha_{\rm halo}\big|
\big|v_m g(v_m)- G(v_m) \big| \,. 
\eeq 
Note that the minus sign between the
two terms is conserved, while in the general case, eq.~\eqref{eq:M1}, one is
forced to sum the two terms in the bound. From eqs.~\eqref{eq:bound_g} 
and \eqref{eq:bound_G_impr} one then obtains the following bound on the
modulation
\beq \label{eq:M2:special}
A_\eta(v_m) \leq v_e |\sin\alpha_{\rm halo}| 
\max \left[\,\left|\frac{d \overline \eta}{d v_m}\right|, \frac{\overline \eta (v_m)}{v_m}
- \int_{v_m} dv \frac{\overline{\eta}(v)}{v^2}
\right] \,.
\eeq
A similar bound has been obtained recently in \cite{Frandsen:2011gi}. If
eq.~(B4) of \cite{Frandsen:2011gi} is expanded in $u$ our bound
\eqref{eq:M2:special} is obtained, if the derivative term dominates.
However, eq.~(B4) of \cite{Frandsen:2011gi} seems to neglect the possibility
that the second term in eq.~\eqref{eq:M2:special} dominates. This assumption is justified if data show no phase flip in the modulation and correspond to $v_m$ above the last phase flip ($A'_\eta \ge 0$). 
In this case eq.~\eqref{eq:M2:special} becomes
\beq \label{eq:M2:special2}
A'_\eta(v_m) \leq - v_e \sin\alpha_{\rm halo}
\frac{d \overline \eta}{d v_m} \,.
\eeq
Note that this bound is complementary to the ones from 
eqs.~\eqref{eq:bound_spec}, \eqref{eq:bound_spec_alpha}. By taking the
integral over the modulation amplitude those bounds probe global properties
of the amplitude over the considered energy range, whereas
eq.~\eqref{eq:M2:special2} bounds the local size of the modulation
amplitude at a given value of $v_m$. Both bounds are necessary conditions
which a modulation signal with a DM origin has to fulfill.

\section{Applying the bound to data}
\label{sec:data}

In this section we show how the bounds \eqref{eq:bound_gen},
\eqref{eq:bound_spec}, \eqref{eq:bound_spec_alpha} obtained in $v_m$ space
for halo integrals can be applied to observable quantities, the unmodulated
rate $\overline R$ and the amplitude of the modulation of the rate, $A_R$. The halo integral
and the DM scattering rate are proportional to each other, see
eq.~\eqref{eq:Rgamma}. The relation is complicated by the fact that
experiments typically do not observe the recoil energy $E_{nr}$ directly.
The nuclear recoil energy is related to the observed energy through a
quenching factor $Q$. In the case of CoGeNT and DAMA the observed energy
$E_{ee}$ is measured in electron equivalent and is related to the recoil
energy by $Q = E_{ee}/E_{nr}$. In general this is a nonlinear equation, as 
$Q$ also depends on the recoil energy. Finally, in an experiment data is
reported in bins, and the continuous bounds derived above have to be
integrated over the bin sizes. We relegate the details of the derivation to
appendix~\ref{sec:translation}, and quote below only the final results.

\subsection{Single target detector}

Let us first assume that the target consists of a single material, as is for
instance the case for CoGeNT. We denote the modulation amplitude and the
unmodulated rate in bin $i$ as $A_i$ and $R_i$, respectively, both in units
of counts/day/kg/keV$_{ee}$.  The general bound \eqref{eq:bound_gen} then
becomes
\begin{align}
A_i 
 &\leq v_e \left[
R_i(\alpha_i +\beta_i) - R_{i+1} \alpha'_{i+1}
 - \ev{\kappa}_i \sum_{j=i}^N R_j \gamma_j \right] 
\qquad\text{(Assumption 1)}\,, \label{eq:general_binned}
\end{align}
and the bounds for the symmetric halo are
\beq
\sum_{j=i}^N A_j x_j \leq v_e \sin\alpha 
\left[
R_i y_i - \ev{v_m}_i \sum_{j=i+1}^N R_j \gamma_j 
\right]
\qquad\text{(Assumptions 2, 2a)}\,, \label{eq:symm_binned}
\eeq
where the bounds \eqref{eq:bound_spec} (Assumption~2) and
\eqref{eq:bound_spec_alpha} (Assumption~2a) are obtained for $\sin\alpha =
1$ and 0.5, respectively. The bin index $i$ runs from 1 to $N$, while the
rates $R_i$ are zero for $i > N$. The coefficients $\alpha_i, \alpha'_i, \beta_i,
\ev{\kappa}_i, \gamma_i, x_i, y_i, \ev{v_m}_i$ are given in
eqs.~\eqref{eq:def_consts1} and \eqref{eq:def_consts2}. They are known
quantities calculable in terms of the form factor $F(E_{nr})$, quenching
factor $Q(E_{nr})$, and Jacobians needed for changing the variable from $v_m$ to $E_{ee}$.
They depend on the DM mass $m_\chi$ via the reduced mass $\mu_{\chi A}$,
eq.~\eqref{eq:vm}, needed to convert $v_m$ into $E_{nr}$. The dependence on
the scattering cross section and the local DM density is encoded in the
constant factor $C$, eq.~\eqref{eq:Rgamma}. This factor cancels completely, as
expected, since it is a common factor for the modulation as well as for the rate.

The rates $R_i$ in the bounds \eqref{eq:general_binned} and
\eqref{eq:symm_binned} are the unmodulated scattering rate
induced by DM without including any backgrounds. In a ``background free''
experiment, where the full observed event rate is due to DM,
the bounds can be  applied as they are. Here we want to be more conservative,
and consider also the situation where an unknown background may contribute
to the unmodulated rate. The remaining assumption is then just that the background itself is not
modulated, only the DM signal. In each bin a fraction $\omega_i$ of the
observed count rate $\mathcal{R}_i$ is due to DM, i.e.,
\beq
R_i = \mathcal{R}_i \omega_i \,,\qquad 0\le\omega_i\le 1\,,
\eeq
so that we can replace $R_i \to \mathcal{R}_i \omega_i$ in
eqs.~\eqref{eq:general_binned} and \eqref{eq:symm_binned} to obtain,
\begin{align}
&A_i 
 \leq v_e \left[
\mathcal{R}_i \omega_i (\alpha_i +\beta_i) - 
\mathcal{R}_{i+1} \omega_{i+1} \alpha'_{i+1}
 - \ev{\kappa}_i \sum_{j=i}^N \mathcal{R}_j \omega_j \gamma_j \right] 
&\text{(Assumption 1)}\,, \label{eq:gen_bin1} \\
&\sum_{j=i}^N A_j x_j \leq v_e \sin\alpha 
\left[
\mathcal{R}_i \omega_i y_i - \ev{v_m}_i \sum_{j=i+1}^N \mathcal{R}_j \omega_j \gamma_j 
\right]
&\text{(Assumptions 2, 2a)}\,. \label{eq:sym_bin1}
\end{align}
Now we have to find a set of $\omega_i$ ($i=1,{\ldots},N$), such that the
bound becomes the ``weakest''. In the following we describe two different
procedures for this task. Procedure~1 is easy to implement but gives
slightly weaker bounds, whereas procedure~2 involves an optimization
algorithm but gives somewhat more stringent bounds.

\bigskip
{\bf Procedure 1:} We can sum eq.~\eqref{eq:gen_bin1} from bin $i$ to
$N$ and drop the last term in the square bracket, since $\ev{\kappa}_i$ and
$\gamma_i$ are positive:
\begin{align}
\sum_{j=i}^N A_j
 &\leq v_e \left[
 \mathcal{R}_i \omega_i \alpha_i + 
 \sum_{j=i}^N \mathcal{R}_j \omega_j \beta_j \right] 
 \leq v_e \left[
 \mathcal{R}_i \alpha_i + 
 \sum_{j=i}^N \mathcal{R}_j \beta_j \right] 
 \qquad \text{(Assumption 1)}\,, \label{eq:proc1general}
\end{align} 
where the last inequality was obtained by setting $\omega_i = 1$. Similarly
we can drop the second term from eq.~\eqref{eq:sym_bin1} and obtain
\begin{align}
\sum_{j=i}^N A_j x_j 
\leq v_e \sin\alpha \, \mathcal{R}_i \omega_i y_i 
\leq v_e \sin\alpha \, \mathcal{R}_i y_i
\qquad \text{(Assumptions 2, 2a)}\,, \label{eq:sym_bin2}
\end{align}
where the last inequality holds for $y_i > 0$.\footnote{According to
eq.~\eqref{eq:def_consts2} $y_i$ can also become negative. In that case the
bound \eqref{eq:sym_bin2} is always violated for any $\omega_i \ge 0$.}
These bounds have to be satisfied for all bins $i$.

\bigskip 

{\bf Procedure 2:} A somewhat stronger bound can be obtained by searching
for the optimal choice for the $\omega_i$, without dropping the last terms
in eqs.~\eqref{eq:gen_bin1} and \eqref{eq:sym_bin1}. We present here a
method based on a least-square minimization (an alternative
procedure is outlined in appendix~\ref{app:proc2}). Let us denote the
r.h.s.\ of eqs.~\eqref{eq:gen_bin1} and \eqref{eq:sym_bin1} as $B_i$ which
are functions of $\omega_j$. Then we can construct from
eq.~\eqref{eq:gen_bin1} the following least-square function
\beq\label{eq:chisq_gen}
X^2 = \sum_{i=1}^N 
\left(\frac{A_i - B_i}{\sigma^A_i}\right)^2
\Theta(A_i - B_i)  \qquad \text{(Assumption 1)} \,,
\eeq
where $\sigma^A_i$ is the $1\sigma$ error on $A_i$ (errors on $B_i$ are
typically much smaller and we neglect them here). This $X^2$ can
now be minimized with respect to $\omega_j$ under the condition $0\le
\omega_j \le 1$. The $\Theta$ function takes into account that there is only
a contribution to $X^2$ if the bound is violated. Hence,
$X^2$ will be zero if the bound is satisfied for all bins. A
non-zero value of $X^2$ indicates that the bound is violated for
some bin(s), weighted by the corresponding error in the usual way. 
In the case of eq.~\eqref{eq:sym_bin1} one has to take into account that the
modulation amplitude in each bin is used several times, leading to
correlated errors for the l.h.s.: 
\beq\label{eq:chisq_sym}
X^2 = \sum_{i,j=1}^N 
(\mathcal{A}_i - B_i) S^{-1}_{ij} (\mathcal{A}_j - B_j)
\Theta(\mathcal{A}_i - B_i) \Theta(\mathcal{A}_j - B_j)  
\qquad \text{(Assumption 2,2a)} \,,
\eeq
where 
\beq
\mathcal{A}_i \equiv \sum_{k=i}^N A_k x_k
\qquad\text{and}\qquad
S_{ij} = \sum_{k=1}^N
\frac{d\mathcal{A}_i}{d A_k}
\frac{d\mathcal{A}_j}{d A_k} \left(\sigma^A_k \right)^2
= \sum_{k={\rm max}(i,j)}^N \left(x_k \sigma^A_k \right)^2
\,.
\eeq

While non-zero values of the $X^2$ functions \eqref{eq:chisq_gen} and
\eqref{eq:chisq_sym} can be considered as a qualitative measure for the
violation of the bound, the precise distribution of them should be
determined by Monte Carlo studies. From the definition one can expect,
however, that they will be approximately $\chi^2$ distributed if the bound
is violated.

\subsubsection{CoGeNT}

\begin{table}
\begin{tabular}{c|c|c|c|c}
\hline\hline 
$E_{ee}$ bins [keV] & Mod.\ (Ass.\ 1, 2) & Mod.\ (Ass.\ 2a) & 
Unmod.\ rate & Corrected rate \\ 
\hline 
0.5--0.9 & $1.41\pm0.79$ & $0.90\pm 0.72$ & $12.33\pm 0.52$& $5.29\pm 0.52$\\ 
0.9--1.5 & $0.84\pm0.59$ & $0.37\pm 0.55$ & $4.33\pm 0.39$ & $3.36\pm 0.39$\\ 
1.5--2.3 & $0.46\pm0.24$ & $0.48\pm 0.22$ & $2.76\pm 0.16$ & $2.76\pm 0.16$ \\ 
2.3--3.1 & $0.66\pm0.24$ & $0.27\pm 0.23$ & $2.83\pm 0.17$ & $2.83\pm 0.17$\\
\hline\hline 
\end{tabular}
\mycaption{\label{tab:data} CoGeNT data on modulation amplitude
and unmodulated rate~\cite{Aalseth:2011wp} in cnts/day/kg/keV, as reported
in .~6 of ref.~\cite{Fox:2011px}. The modulation has been extracted
allowing for indpendent phases in each bin (Ass.~1) and for a constant but
arbitrary phase (Ass.~2) (which lead to very similar amplitudes), and by
requiring the maximum at June 2nd (Ass.\ 2a). In the last column we show the
preliminary surface events corrected unmodulated rate~\cite{Collar-TAUP}.
}
\end{table}

Let us consider now the modulation signal reported by the CoGeNT experiment
at $2.8\sigma$~\cite{Aalseth:2011wp}. It has been pointed out that for
specific assumptions on the halo (standard Maxwellian) there is a tension
between the modulated and unmodulated rate in CoGeNT~\cite{Frandsen:2011ts,
Schwetz:2011xm, Fox:2011px}. Recent analyses on the CoGeNT modulation signal
can be found in refs.~\cite{Chang:2011eb, Arina:2011zh}, see also
\cite{Frandsen:2011gi}. Here we use the CoGeNT data for a case study and
apply the above bounds assuming spin-independent elastic scattering.  For
the germanium quenching factor we use $E_{ee} [{\rm keV}] = 0.199 (E_{nr}
[{\rm keV}])^{1.12}$~\cite{Barbeau:2007qi}. We adopt the Helm
parameterization of the spin-independent form factor, $F(E_{nr}) = 3 e^{-q^2
s^2/2} [\sin(q r)-q r\cos(q r)] / (q r)^3$, with $q^2 = 2 m_A E_{nr}$ and $s
= 1$~fm, $r = \sqrt{R^2 - 5 s^2}$, $R = 1.2 A^{1/3}$~fm.  We use the data
from fig.~6 of ref.~\cite{Fox:2011px} where the total rate $\mathcal{R}$ and
the modulation amplitude $A_R$ are given in four bins of $E_{ee}$ between
0.5 and 3.1~keV. We reproduce the data in table~\ref{tab:data}.\footnote{
We thank Mariangela Lisanti for providing us the data from fig.~6 of 
ref.~\cite{Fox:2011px}.}

The modulation amplitude has been extracted in three different ways in \cite{Fox:2011px}.
First, by fitting independently the modulation phase for each energy bin,
second, by assuming a constant phase for all bins, and third, by fixing the
phase such that the modulation maximum is on June~2nd. These are precisely
the requirements corresponding to our Assumptions 1, 2, 2a, respectively. We
can thus use the appropriate data on the modulation for each of the three
assumptions. The modulation amplitudes in the first and the second case are
very similar, while in the third case (forcing the phase to equal June~2nd)
the amplitudes are lower and error bars are larger.

\begin{figure}
  \includegraphics[width=0.6\textwidth]{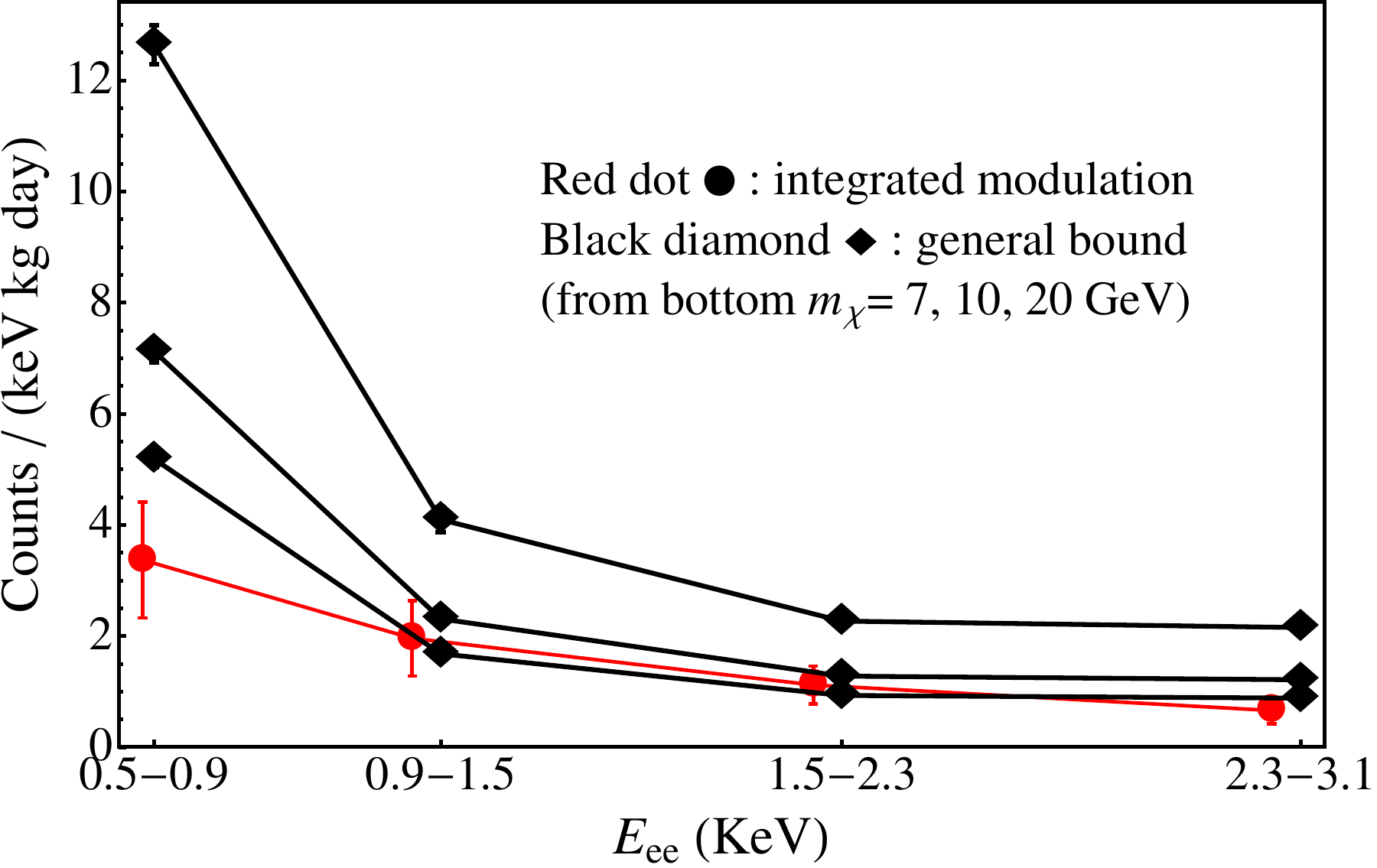}
 
  \mycaption{\label{fig:bounds0} Procedure $1$ upper bound compared to the
  integrated modulation amplitude from the CoGeNT data, from fig.~6 of
  \cite{Fox:2011px}. The red dots correspond to the l.h.s.\ of
  eq.~\eqref{eq:proc1general}. The black diamonds correspond to the upper
  bound obtained under Assumption 1, r.h.s of eq.~\eqref{eq:proc1general},
  for DM masses of $m_\chi = 7, 10, 20$~GeV (from bottom to top) . The bins
  on the horizontal axis indicate the bin $i$ from which we start to sum
  the data.  Error bars correspond to $1\sigma$.}
\end{figure}

Choosing three values of the DM mass as examples, we show in
fig.~\ref{fig:bounds0} the bound on the integrated modulation amplitude for
Assumption~1, and using procedure~1. The bin labels on the horizontal axis
give the $i$th energy bin, which is the lower limit of summation in
eq.~\eqref{eq:proc1general}. The data on amplitudes are shown as red dots,
while the bounds are shown in black. Whenever a red dot lies above one of
the bounds (within errors), the DM hypothesis is disfavoured. In
fig.~\ref{fig:bounds0} this happens for the case of DM mass $m_\chi =
7$~GeV.  Such light DM thus cannot be the source of the modulation, even
under the very general assumption on the DM halo adopted here.

\begin{figure}
 \includegraphics[width=0.75\textwidth]{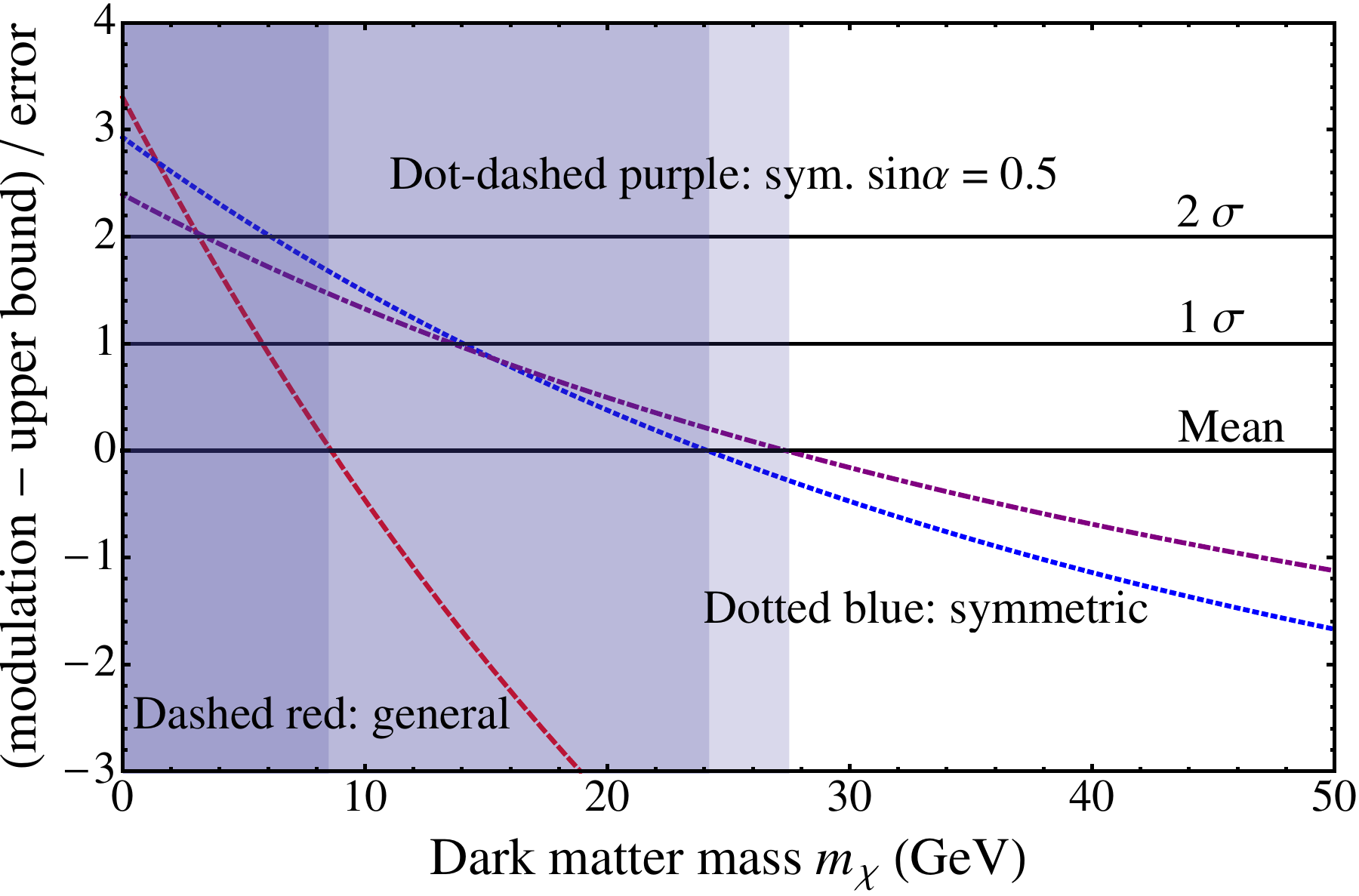}
  \mycaption{\label{fig:bounds} Bounds on the CoGeNT modulation amplitude
  for the three assumptions about the DM halo (1, 2, 2a), and using
  eqs.~\eqref{eq:proc1general} and \eqref{eq:sym_bin2} (procedure 1). We
  show the l.h.s.\ (integrated modulation amplitude) minus the r.h.s.\
  (upper bound) divided by the error on the l.h.s., as a function of DM
  mass. We sum the data starting from bin $i = 3, 2$ and $3$ for Assumptions
  1, 2 and 2a, respectively. The bounds are violated in the regions where
  the curves are above zero, which are shaded in the plot.}
\end{figure}

We see from fig.~\ref{fig:bounds0} that the strongest restriction comes from
the bins $i = 2$ or 3. In fig.~\ref{fig:bounds} we show the constraints on
DM mass that follow from the bounds on the modulation amplitudes in these
two bins. The bounds are obtained from procedure~1 for the three Assumptions
1, 2, 2a. We observe that the bound becomes stronger for smaller DM masses.
This behaviour can be understood from how the coefficients defined in
appendix~\ref{sec:translation} depend on the reduced mass $\mu_{\chi A}$.
The bounds on amplitudes are violated for DM masses below 10, 27, 33~GeV,
respectively, for the three assumptions. The lower bounds on $m_\chi$ are
summarized in table~\ref{tab:bounds} (left part), where we also give the
bounds at 1$\sigma$ and 2$\sigma$. Notice that due to the smaller modulation
amplitudes and larger errors when extracted from the data under
Assumption~2a, at $95\%$~CL the corresponding bound becomes weaker than the
one from Assumption~2 and equal to the one from Assumption~1, although
naively one would expect the opposite.  This can be traced back to the fact
that, under Assumption~2a, the modulation phase is forced to take the value
of June 2nd which is not the one preferred by the data. The extracted
modulation signal then gets weaker and consequently the bounds are more
easily satisfied. We have also checked that bounds for
Assumptions 2, 2a derived from eq.~\eqref{eq:M2:special2} give always weaker
limits than the ones discussed here, which are based on eq.~\eqref{eq:bound_spec} and
eq.~\eqref{eq:bound_spec_alpha}.

\begin{table}
\begin{tabular}{ll|c|c|c|c|c|c|c|c}
\hline\hline
& & \multicolumn{4}{|c|}{unmodulated rate from \cite{Aalseth:2011wp}} &
    \multicolumn{4}{|c }{corrected unmod.\ rate \cite{Collar-TAUP}} \\
\hline    
& & {Mean} & {68\%} & {95\%} & $X^2 \le 1$
  & {Mean} & {68\%} & {95\%} & $X^2 \le 1$\\
\hline
Ass.\ 1:  & general bound  & 8.5 &  6 &   3 & 7.3 &
                             10 &  6.5 & 3 & 10  \\
Ass.\ 2:  & symmetric halo & 24 & 14 &  6   & 18 &
                             43 & 25 & 12.5 & 37 \\
Ass.\ 2a: & sym.\ halo, $\sin\alpha = 0.5$ & 
27.5 & 13.5  & 3.5 & 16 &
59.5 & 23  & 3 & 35 \\
\hline\hline
\end{tabular}
  \mycaption{\label{tab:bounds} Lower bounds on the DM mass in GeV, from the
  requirement that the modulation amplitude in CoGeNT is consistent with the
  upper bound from the unmodulated rate, according to the Assumptions 1, 2,
  2a on the DM distribution. The bounds are obtained from procedure 1,
  requiring that eqs.~\eqref{eq:proc1general} or \eqref{eq:sym_bin2} are
  satisfied for the mean value, or the 68\% and 95\%~CL limits. The bounds
  labeled ``$X^2 \le 1$'' are obtained from procedure~2 by
  requiring that $X^2$ defined in eqs.~\eqref{eq:chisq_gen} or
  \eqref{eq:chisq_sym} is less than 1. In the left part of the table we use
  the published unmodulated event rates from  \cite{Aalseth:2011wp}, whereas
  for the right part of the table we adopt the preliminary results on  
  surface events contamination at low energies from \cite{Collar-TAUP}.}
\end{table}

\begin{figure}
 \includegraphics[width=0.7\textwidth]{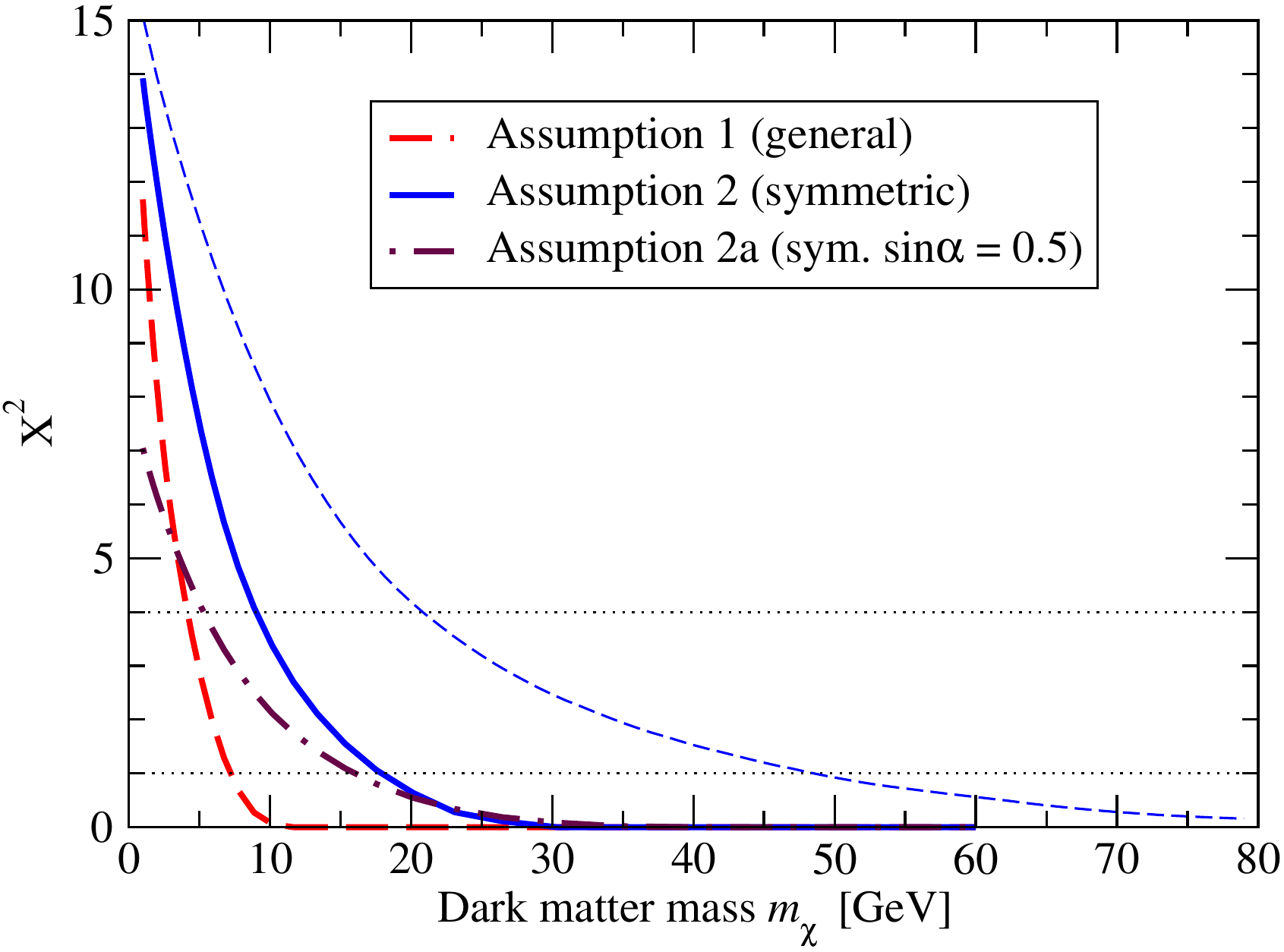}
  \mycaption{\label{fig:chisq} Bounds on the CoGeNT modulation amplitude for
  the three assumptions about the DM halo (1, 2, 2a), always using procedure
  2. We show $X^2$ defined in eqs.~\eqref{eq:chisq_gen} and
  \eqref{eq:chisq_sym} as a function of DM mass. The thin dashed curve is
  for illustrative purpose only; it corresponds to the data and errors on
  the modulation amplitude extracted with a free phase (Assumption 2) but
  using the bound for Assumption 2a, see text for details. If we assume that
  $X^2$ is distributed as a $\chi^2$ with 1 d.o.f.\ (see below) $X^2 =
  1\,(4)$ corresponds to 68\% (95\%) CL.} 
\end{figure}

In fig.~\ref{fig:chisq} we show the $X^2$ functions according to
procedure~2 as a function of the DM mass. For a given value of $m_\chi$ we
minimize $X^2$ numerically with respect to the $\omega_j$. The
conclusion is similar to procedure 1, leading to similar lower bounds on the
DM mass. Requiring that $X^2 \le 1$ one finds $m_\chi \ge 7.3, 18,
16$ GeV for Assumptions~1, 2, 2a, respectively. We observe again the unusual
situation that Assumption~2a leads to a weaker constraint, due to the less
significant signal for the modulation. To illustrate this effect we show in
fig.~\ref{fig:chisq} with a thin-dashed curve also $X^2$ that would
follow from the hypothetical situation where the modulation amplitude would
be as strong as in the case of Assumption~2. That is, we allow for an
arbitrary energy independent phase, but assume that this phase turned out to
be June 2nd (in contrast to the real CoGeNT data), so that we could apply
the bound from Assumption~2a to these data. We see that in this case one
would obtain a much stronger bound, disfavoring DM masses up to 60~GeV. This
example shows the potential of our method in cases of a strong modulation
signal in the data. If the signal for the modulation itself is weak, the
bounds we derived will also give only weak constraints.

\begin{figure}
  \includegraphics[height=0.4\textwidth]{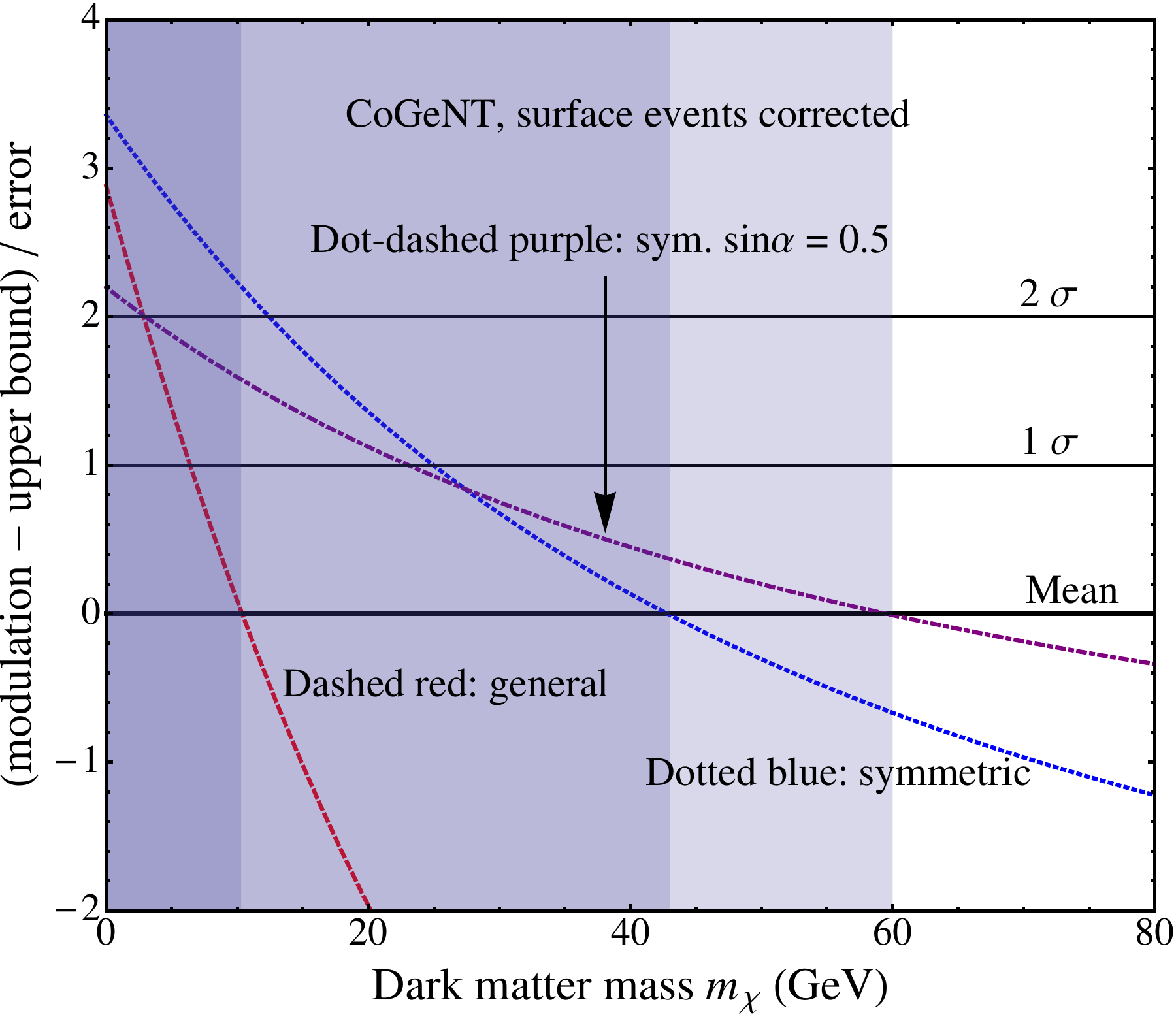}
  \includegraphics[height=0.4\textwidth]{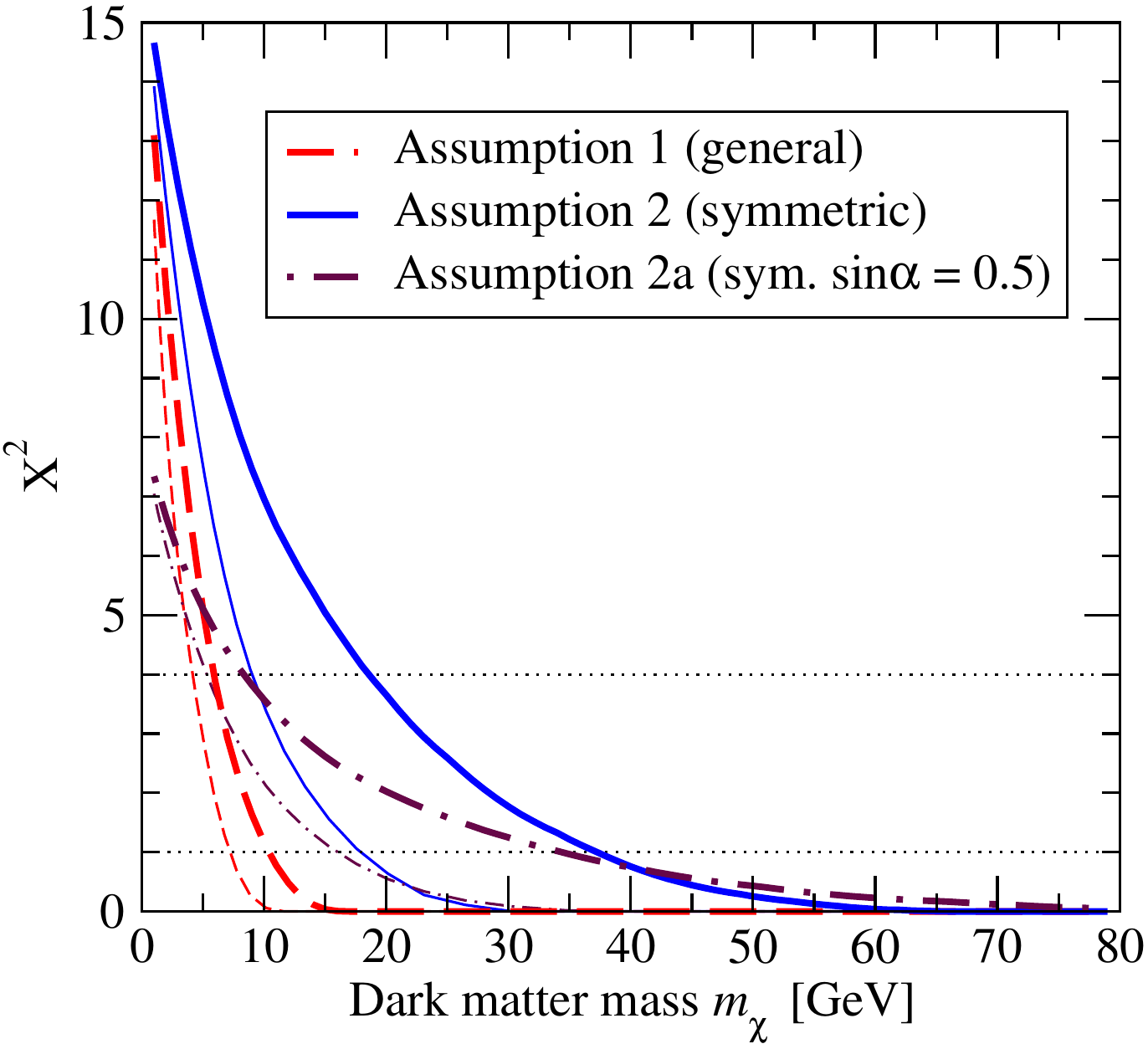}
  \mycaption{\label{fig:surface-ev} Bounds on the CoGeNT modulation
  amplitude for the preliminary surface events corrected unmodulated
  rate~\cite{Collar-TAUP} for the three assumptions about the DM halo (1, 2,
  2a). The left plot shows the bounds according to procedure~1, whereas the
  right plot shows $X^2$ defined in eqs.~\eqref{eq:chisq_gen} and
  \eqref{eq:chisq_sym} (procedure~2). The thin curves in the right panel are
  the bounds without surface event subtruction reproduced from
  fig.~\ref{fig:chisq} for the purpose of comparison with the thick curves
  obtain with the surface event corrected rates.}
\end{figure}

A recent re-analysis of CoGeNT data indicates that a significant fraction of
the event excess at low energies could be due to surface
events~\cite{Collar-TAUP}. Assuming that surface events are not modulated,
the unmodulated rate will be reduced after subtructing the surface events,
while the modulation signal will remain, leading to a strengthening of our
bounds. Here we estimate this effect by using the preliminary result for a
surface events rejection efficiency shown on slide~19 of the presentation
in~\cite{Collar-TAUP} (red curve). Averaging this curve for the bins used in
our analysis we find that the unmodulated rate in the first and second bins
are reduced by a factor 0.43 and 0.78, respectively, while the other two
bins are not effected. This leads to the reduced event rates shown in the
last column of tab.~\ref{tab:data}.  Fig.~\ref{fig:surface-ev} shows the
plots equivalent to figs.~\ref{fig:bounds} and \ref{fig:chisq} but using the
surface events corrected unmodulated rate. The general bound remains
essentially the same. In this case the strongest limit (from procedure~1)
for the uncorrected rate comes from summing data starting from bin~3,
whereas in the surface events corrected case it comes from bin~2, with only
a minor reduction (79\%) of the rate, leading to a very similar limit.  In
contrast, for Assumptions~2 and 2a significantly stronger bounds are
obtained, with the limit coming now from summing from bin~1, with a 43\%
reduced rate. The surface events corrected bounds are summarized in the
right part of table.~\ref{tab:bounds}.

Due to the non-standard definition of the $X^2$ functions
\eqref{eq:chisq_gen} and \eqref{eq:chisq_sym}, involving the
$\Theta$-function, the actual distribution of them is not clear a priori.
Therefore we have performed a Monte Carlo study in order to determine the
distribution. For a given DM mass we first determine the optimal set of
$\omega_i$ by minimizing the $X^2$. For those $\omega_i$ we assume that
the bound is saturated and we simulate a large number of pseudo-data taking
the r.h.s.\ of eqs.~\eqref{eq:gen_bin1}, \eqref{eq:sym_bin1} as mean value
for a Gaussian with standard deviation given by the actual error on the
modulation amplitude.\footnote{In the case of Assumptions 2 and 2a we
proceed iteratively. Starting from eq.~\eqref{eq:sym_bin1} for $i = N$ we
generate $A_N$, and then simulate successively $A_{i-1}$. This is necessary
to obtain the correct random properties of the l.h.s.\ of
eq.~\eqref{eq:sym_bin1}.} For each random data set we calculate the $X^2$
value and obtain therefore the distribution of $X^2$ assuming that the
bound is saturated. Then we can compare the $X^2$ of the actual data and
calculate the probability of obtaining a $X^2$ larger than the observed
one. Note that this procedure is conservative, since we assume that the mean 
value of the random data is the bound itself. If the ``true'' mean values
are smaller than the bound, the distribution of $X^2$ would be shifted
towards smaller values, leading to more stringent bounds. 

\begin{figure}
  \includegraphics[width=0.6\textwidth]{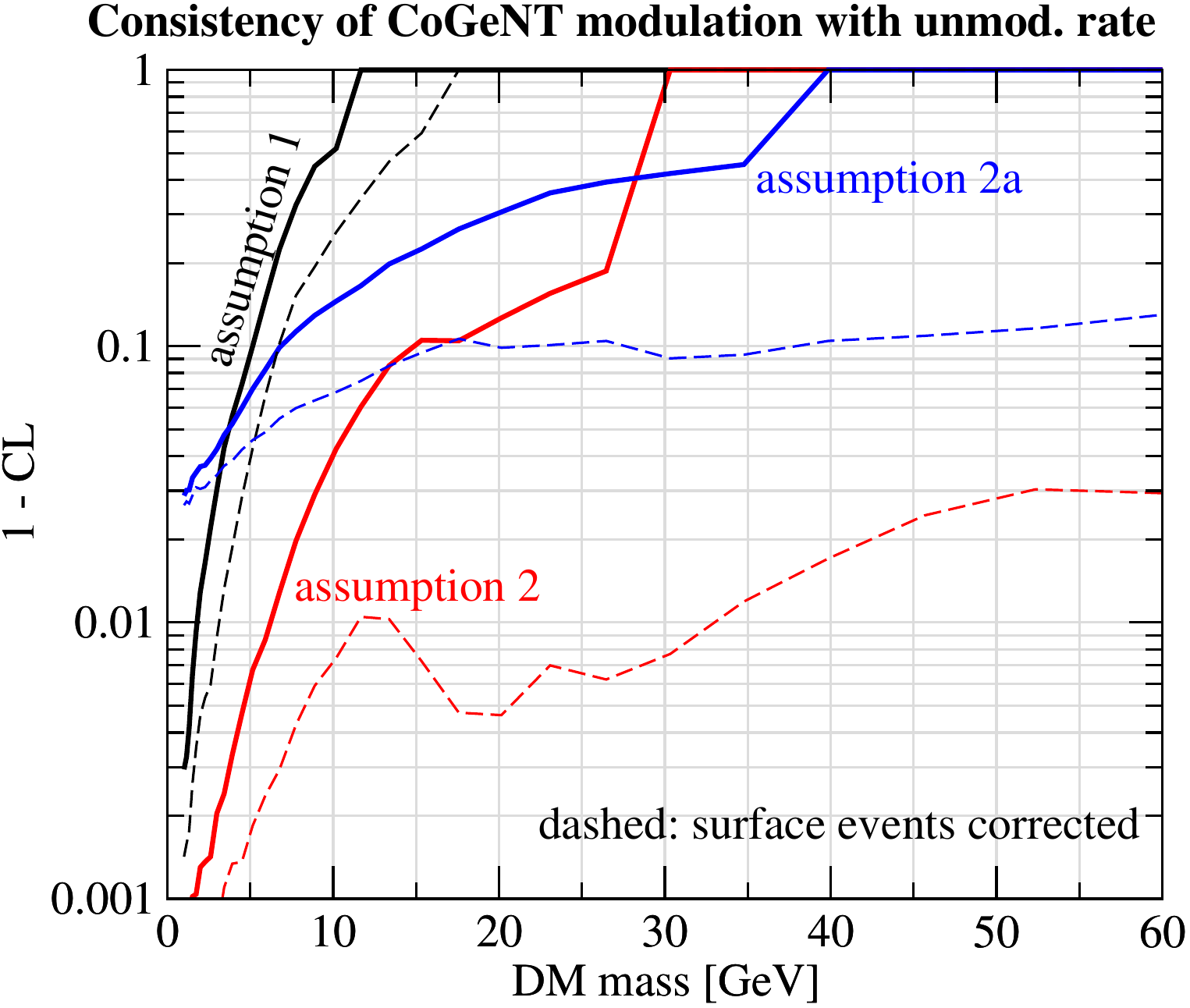}
  \mycaption{\label{fig:prob} The probability to obtain a $X^2$ larger than
  the one obtained from CoGeNT data, as determined by Monte Carlo simulation.
  Dashed curves correspond to surface event corrected data.}
\end{figure}

We show the results of this calculation in fig.~\ref{fig:prob}. The
probabilities obtained in this way are in qualitative agreement with the
numbers reported in tab.~\ref{tab:bounds}. Under assumptions 2 and 2a,
surface event corrected data is inconsistent with the DM hypothesis for any
DM mass at the 97\% and 90\%~CL, respectively. We find that if the bound is
violated (i.e., for small DM masses) the $X^2$ distribution is close to a
$\chi^2$-distribution with 1 d.o.f.. Large deviations from a
$\chi^2$-distribution occur if $X^2$ is close to zero (i.e., for larger DM
masses). In this case the $X^2$ distribution is strongly peaked at small
values close to zero. This indicates that in such situations it is in most
cases possible to find a set of $\omega_i$ such that $X^2$ becomes zero.
This is the reason why non-trivial constraints are obtained under
assumptions 2, 2a for surface events corrected data at large masses,
although the values of $X^2$ are relatively small, compare
fig.~\ref{fig:surface-ev}.

\subsection{Bounds for multi-target experiments}

So far we have restricted the discussion to the situation where only
one target nucleus is present, as for example Ge in CoGeNT. Let us now
generalize our bounds to experiments where several elements are used
as DM target. In this case both the modulation amplitude as well as
the unmodulated rate will receive contributions from each element:
$A_i=\sum_n A_i^n$ and $R_i=\sum_n R_i^n$, where $n$ labels the
different target elements and $i$ energy bins. The bounds from
procedure~1 hold for each of the elements separately, where in general
all the coefficients appearing in the equations will depend on the
nucleus type. Summing eqs.~\eqref{eq:proc1general} and
\eqref{eq:sym_bin2} over $n$ and using the fact that for positive
$a_i$ and $b_i$, the following inequality holds, $\sum_i a_i b_i \leq
(\sum_i a_i)(\sum_j b_j)$, we derive: 
\begin{align}
&\sum_{j=i}^N A_j
  \leq v_e \left[
 \mathcal{R}_i  \sum_n \alpha_i^n  + 
   \sum_{j=i}^N \mathcal{R}_j \sum_n | \beta_j^n | \right] 
   \qquad\text{(Assumption 1)}\,, \label{eq:proc1generalb} \\
&\sum_{j=i}^N A_j \, {\rm min}_n \left( x_{j}^n \right)
  \leq 
  v_e \sin\alpha \, \mathcal{R}_i \sum_n | y_i^n| 
  \qquad \text{(Assumptions 2, 2a)}\,. \label{eq:sym_bin2b}
\end{align}
On the l.h.s.\ of the last inequality it is necessary to take the minimum 
of $x_{j}^n$ between all the nuclei present in the target, for each bin. We
used that $\alpha_i$ and $x_j$ are positive, see
appendix~\ref{sec:translation}.

It is possible to generalize also the $X^2$ method to the
case of multiple targets. For each nucleus the bounds
\eqref{eq:gen_bin1} and \eqref{eq:sym_bin1} apply separately. We
define the amplitude from element $n$ in the bin $i$ as $A_i^n =
\epsilon^n_i A_i$, and similarly for the rate $R_i^n = \omega^n_i
\mathcal{R}_i$.  Then one can construct a $X^2$ similar to the one in
the previous section, from l.h.s.\ minus r.h.s.\ of the following inequality:
\begin{align}
&\epsilon^n_i A_{i} 
 \leq v_e \left[
\mathcal{R}_{i} \omega^n_i (\alpha^n_i +\beta^n_i) - 
\mathcal{R}_{i+1} \omega^n_{i+1} \alpha^n_{i+1}
 - \ev{\kappa^n}_i \sum_{j=i}^N \mathcal{R}_{j} \omega^n_j \gamma^n_j \right]
&\quad\text{(Assumption 1)}\,. \label{eq:gen_bin1_multi} 
\end{align}
This $X^2$ is minimized over $\omega^n_i \ge 0$, $\epsilon^n_i \ge
0$ given the constraints $\sum_n \omega^n_i \leq 1$ and $\sum_n
\epsilon^n_i=1$ for each $i$. A similar relation can be set up for
Assumption~2.

\subsubsection{DAMA}

\begin{figure}
 \includegraphics[width=0.48\textwidth]{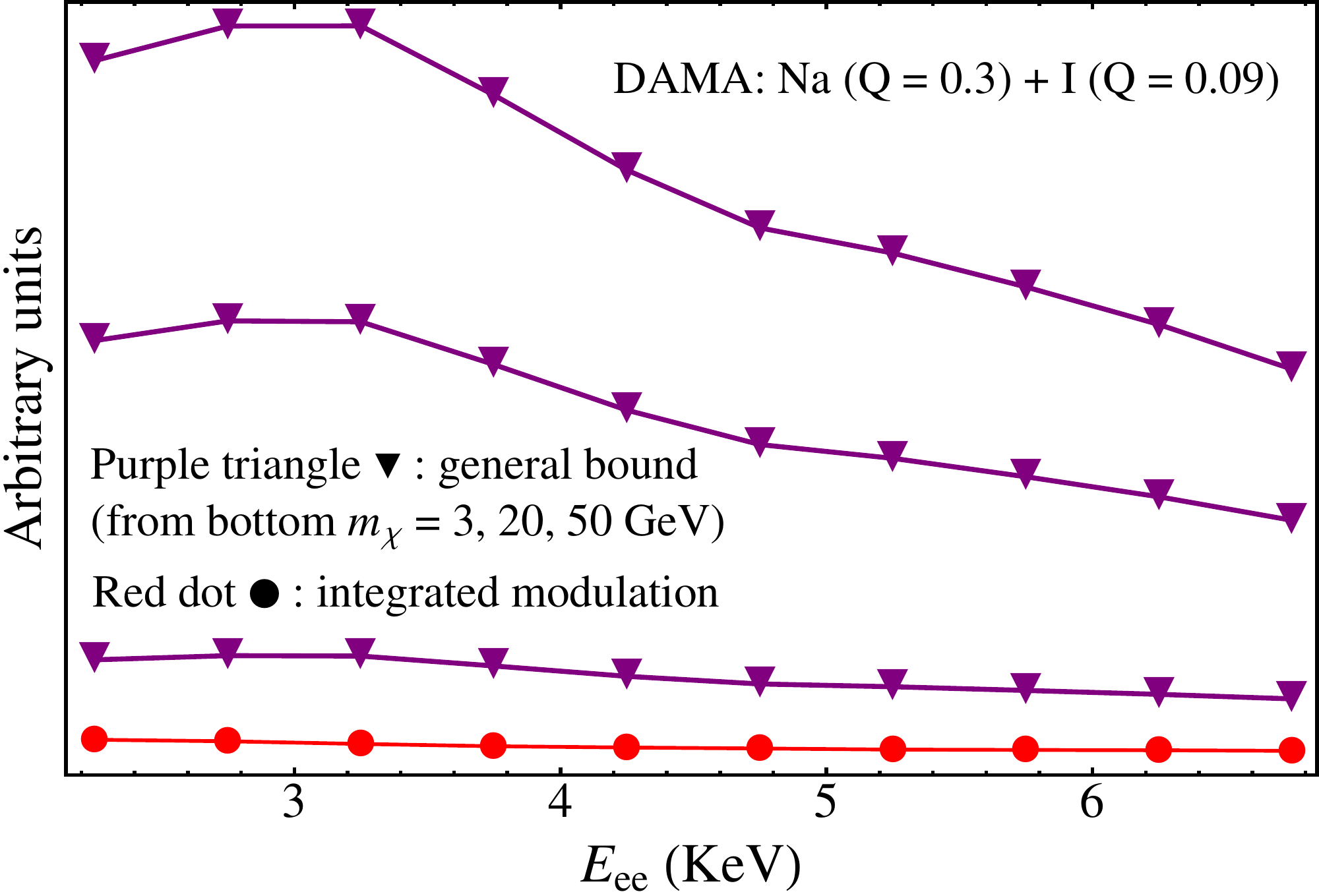}
 \includegraphics[width=0.48\textwidth]{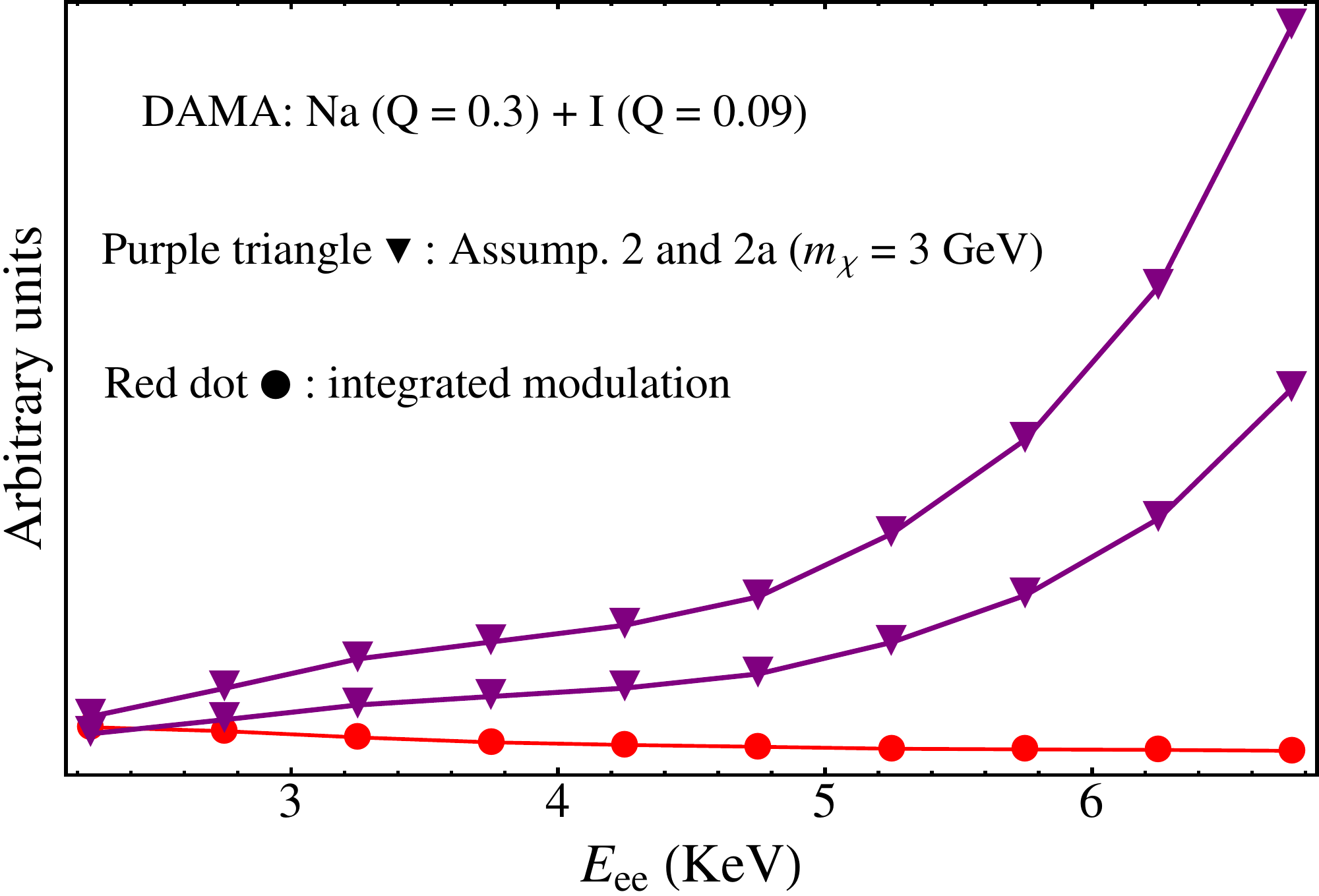}
 \mycaption{\label{fig:DAMA} Upper bound compared to
   the integrated modulation amplitude for DAMA data 
   for Assumption~1 (left) and Assumptions 2 and 2a (right). 
   We assume quenching factors $q=0.3$ and
   $q=0.09$ for Na and I, respectively. The red dots correspond to the
   l.h.s.\ of eqs.~\eqref{eq:proc1generalb} and \eqref{eq:sym_bin2b},   
   and the purple triangles to
   the r.h.s.. In the left panel we show the bounds    
   for DM masses of $m_\chi = 3, 20, 50$~GeV (from
   bottom), in the right panel for $m_\chi = 3$~GeV.   
   Bins are $0.5$~keV wide and we sum all bins starting from
   bin $i$ shown on the horizontal axis up to $7$~keVee. Error bars
   are negligible and are not shown for clarity. All dark matter
   masses are compatible with the modulation.}
\end{figure}

Let us now apply above bounds to the DAMA result, based on data from a NaI
detector. DAMA observes a relatively large unmodulated count rate of about
1~cnt/kg/day/keV compared to the modulation amplitude of about
0.02~cnt/kg/day/keV. In contrast, CoGeNT reports a modulation amplitude in
the range of 0.5 to 1~cnt/kg/day/keV, compared to an unmodulated rate of 3
to 4~cnt/kg/day/keV. Because of this much smaller modulation amplitude
compared to the unmodulated rate for DAMA we expect the signal to be
consistent with our bounds. It is also well known that a consistent fit of
the DAMA modulated and unmodulated data is possible e.g., for a Maxwellian
halo, although in certain parameter regions the unmodulated rate does
provide a non-trivial constraint, see e.g.~\cite{Fairbairn:2008gz, Chang:2008xa}.  In
fig.~\ref{fig:DAMA} we show the bounds for DAMA, assuming quenching factors
of $0.3$ for Na and $0.09$ for I (constant in energy) using the spectral
data on the modulated and unmodulated rates from~\cite{Bernabei:2008yi}. We
find that both the general bound, eq.~\eqref{eq:proc1generalb}, as well as
the bounds for a symmetric halo, eq.~\eqref{eq:sym_bin2b}, are consistent
with the modulation even for a DM mass as small as 3~GeV. As noted above the
bounds get weaker for larger DM masses, hence they do not place any relevant
constraint on the dark matter interpretation of the DAMA modulation signal.

\begin{figure}
 \includegraphics[width=0.48\textwidth]{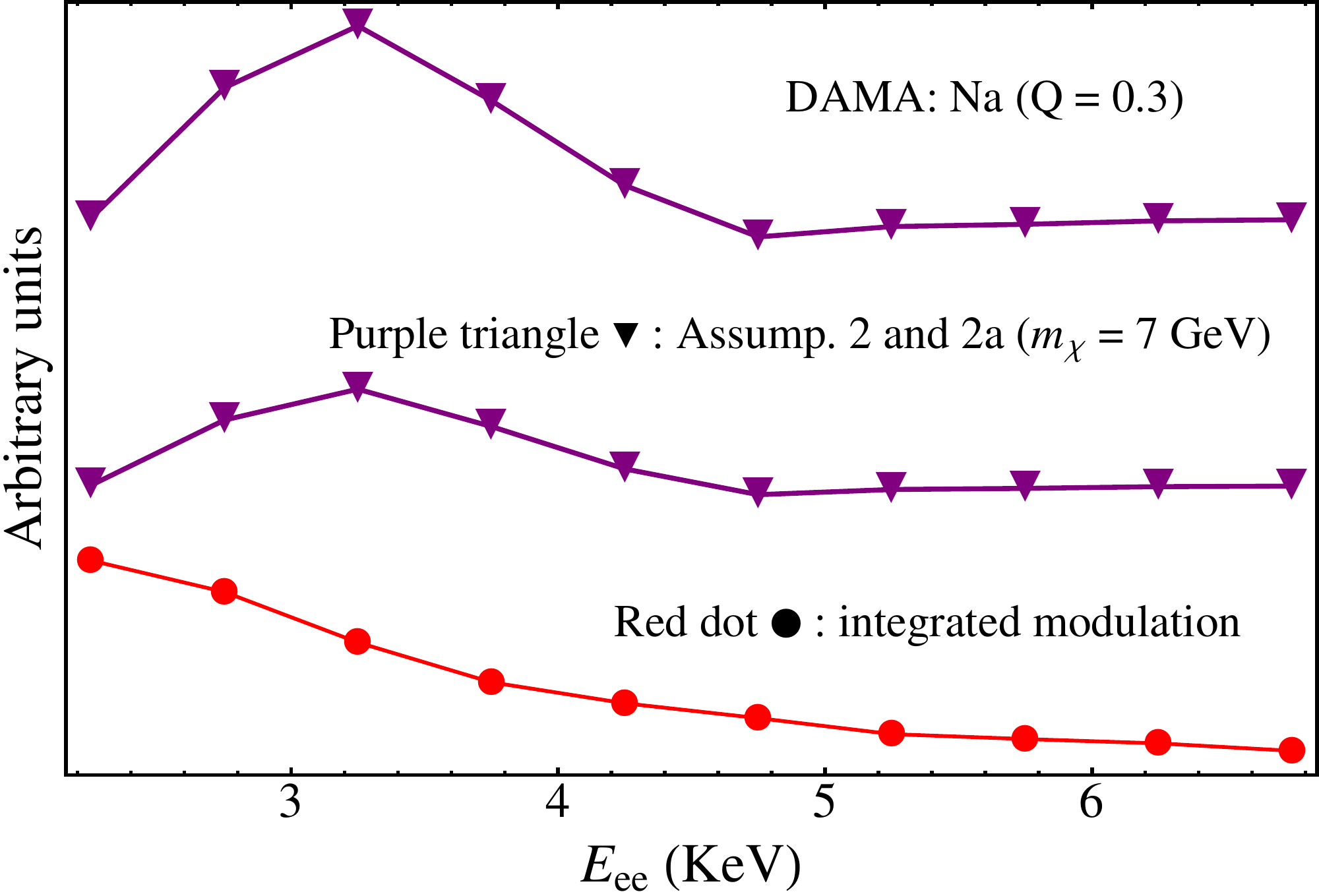}
 \includegraphics[width=0.48\textwidth]{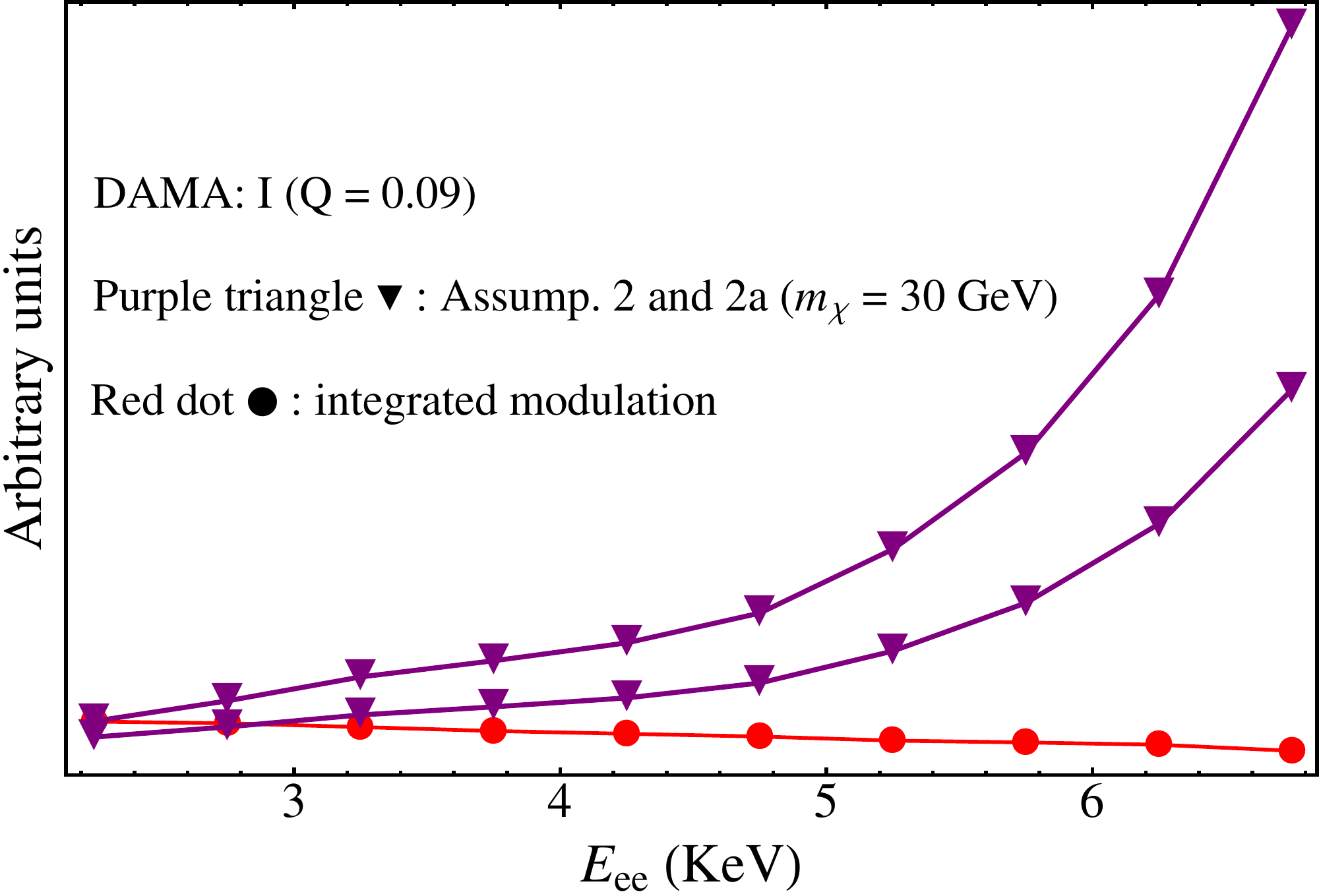}
 \mycaption{\label{fig:DAMA3} Upper bound compared to the integrated
   modulation amplitude from DAMA data for Assumptions $2$ and $2$a,
   assuming that DM scatters only on sodium (left) or
   iodine (right). The red dots correspond to the l.h.s.\ of
   eq.~\eqref{eq:sym_bin2} and the purple triangles to the r.h.s., for
   a DM mass of $m_\chi = 7$~GeV (left) and 30~GeV (right). Bins are
   $0.5$~keV wide and we sum all bins starting from bin $i$ shown on
   the horizontal axis up to $7$~keVee. 
   Error bars are negligible
   and are not shown for clarity.}
\end{figure}

In deriving the multi-target bounds eqs.~\eqref{eq:proc1generalb} and
\eqref{eq:sym_bin2b} we had to use inequalities related to the presence of
several contributions to the rates, which makes the bounds somewhat weaker
than single target bounds. In case of NaI we have a relatively large
hierarchy between the two targets since $A = 23$ for Na and $A = 127$ for I.
Therefore, one may consider also the situation that only one of them
dominates the signal. If scattering on both targets is kinematically
allowed, iodine will dominate because of the $A^2$ enhancement of the
spin-independent scattering cross section (which we will assume here). For
low DM masses, scattering on iodine may be kinematically forbidden since
$v_m$ is larger than the escape velocity of the halo and in that case the
signal is provided only by scattering off Na (this requires additional
assumptions on the escape velocity, which in general are not necessary for
our bounds).  We show the bounds for Assumptions 2 and 2a assuming that
scattering on either Na or I dominates in fig.~\ref{fig:DAMA3} for some
representative DM masses. In both cases only very weak constraints are
obtained. While scattering on Na is consistent down to $m_\chi \approx
3$~GeV for Assumption~2a, scattering on iodine is consistent for $m_\chi
\gtrsim 30$~GeV under Assumption~2. (Note that below around 30~GeV typically
scattering on Na dominates.)

As expected, we conclude that based solely on our 
bounds the DAMA signal is compatible with assuming dark matter
scattering being its origin. We do not expect that this conclusion will
change significantly if bounds from procedure 2 according to
eq.~\eqref{eq:gen_bin1_multi} were applied. Therefore, we limited
ourselves to the discussion based on procedure 1 for DAMA.

\section{Summary and discussion}
\label{sec:conclusion}

The annual modulation is probably the most distinctive signature of dark
matter and is playing (and will certainly play) a central role in revealing
its existence and nature. However, it is not always clear that the
modulation detected by a DM direct detection experiment is caused by the
dark matter particles and not by any other unknown background source. In
this work we have presented a consistency check for the amplitude of the
modulation and the unmodulated count rate, which is a necessary (but not
sufficient) condition for DM being the origin of an observed annual
modulation. 

We have derived upper bounds on the energy integrated modulation amplitude
in terms of the unmodulated rate by expanding the DM velocity integral to
first order in the earth's velocity $v_e \approx 30$~km/s. It holds for a wide
class of particle physics models where the differential scattering cross
section $d\sigma/dE_{nr}$ is proportional to $1/v^2$. Although we have only
focused on elastic scattering the method can also be generalized to the
inelastic case. The important aspect of our work is that our bounds hold for
very general assumptions about the DM velocity distribution $f(\vect{v})$ in
the sun's vicinity. We have presented bounds under the hypothesis of a 
single DM species and the following assumptions on the DM halo:

\begin{itemize}
\item
{\bf Assumption 1, bound in eq.~\eqref{eq:bound_gen}:} We assume that the
only time dependence is induced by the rotation of the earth around the sun.
The halo itself is static on the time scale of months to years and spatially
constant at the scale of the sun--earth distance. Otherwise the DM velocity
distribution can have an arbitrary structure, including, for example,
several streams coming from various directions.  In order to saturate this
bound the halo has to be very peculiar, with rather unrealistic properties.

\item
{\bf Assumption 2, bound in eq.~\eqref{eq:bound_spec}:} In addition to
Assumption~1 we assume that there is just one preferred direction of
the DM velocity distribution in the rest frame of the sun (independent
of the minimal velocity $v_m$ in the halo integral). This requires
certain symmetries of the velocity distribution, which are specified
in eqs.~\eqref{eq:cond2} and \eqref{eq:cond3}. It covers typically
situations where the DM distribution is dominated by one single
component, which may come from an arbitrary (but constant)
direction. Assumption~2 requires that the phase of the modulation is
independent of the recoil energy (up to a phase flip of half a year).

\item
{\bf Assumption 2a, bound in eq.~\eqref{eq:bound_spec_alpha}:} We
require that the preferred direction from Assumption~2 is aligned with
the motion of the sun. This is fulfilled for any isotropic halo, and
also for tri-axial halos up to corrections due to the peculiar
velocity of the sun.  Furthermore, it includes the possibility of
streams parallel to the motion of the sun, such as a possible dark
disk. Assumption~2a requires that the maximum (or minimum) of the
event rate is around June~2nd.
\end{itemize}

The theoretical bounds are obtained in terms of the minimal velocity $v_m$
and have to be related to observable quantities by translating into recoil
energy. We have outlined two possible procedures for this task in
section~\ref{sec:data} for the case of elastic scattering, taking into
account statistical errors and the possibility that an unknown background
may contribute to the unmodulated rate, but that it has no time-dependence. As an example we have applied the proposed consistency checks to the annual modulation signals reported by the
CoGeNT and DAMA experiments.  While DAMA data are compatible with a dark
matter origin for its modulation, severe restrictions on the dark matter
mass can be set for the case of CoGeNT. Applying our bounds we find that the
CoGeNT modulation amplitude can be consistent with the unmodulated rate at
the $68\%$~CL only for DM masses $m_{\chi} \gtrsim 7.3, 18, 16$~GeV for the
Assumptions~1, 2, 2a, respectively. If preliminary results on a possible
surface events contamination of the unmodulated rate at low energies in
CoGeNT are confirmed, those bounds would become even more restrictive.
In this case, CoGeNT modulation data would be inconsistent with
the DM hypothesis under assumptions 2 and 2a at about 97\% and 90\% CL,
respectively. DAMA has a relatively large unmodulated count rate of about
1~cnt/kg/day/keV compared to the modulation amplitude of about
0.02~cnt/kg/day/keV. Therefore, our bounds are not very stringent and the
modulation amplitude is consistent with the unmodulated rate. In contrast,
CoGeNT reports a modulation amplitude in the range of 0.5 to
1~cnt/kg/day/keV, compared to an unmodulated rate of 3 to 4~cnt/kg/day/keV.
Because of this relatively large ratio between modulated and unmodulated
rate our method provides stringent constraints in the case of CoGeNT.
Several comments are in order: 

\bigskip

$(i)$ In deriving the bounds we assume a certain smoothness of the
halo, since we are expanding in $v_e$. Spikes in the velocity distribution
much narrower than 30~km/s are not covered by our procedure.

$(ii)$ Our bounds assume that a modulation signal is present in the
data. If the significance of the modulation is weak, the bounds are
more easily satisfied. This effect is explicitly illustrated in
sec.~\ref{sec:data} by using CoGeNT data under different assumptions
regarding the modulation signal. The method discussed here cannot be
used to establish the presence of a modulation, it can only test
whether a given modulation signal is consistent with the unmodulated
rate.

$(iii)$ In this work we have used the relation between the minimal
velocity $v_m$ and nuclear recoil energy $E_{nr}$ implied by elastic
scattering, see eq.~\eqref{eq:vm} and appendix~\ref{sec:translation}.
The bounds can be generalised in a straightforward way also to
inelastic scattering.  However, in that case $v_m = (m_A E_{nr} /
\mu_{\chi A} + \delta)/\sqrt{2m_A E_{nr}}$ has a minimum in $E_{nr}$,
and therefore there is no one-to-one correspondence between $v_m$ and
$E_{nr}$. When translating from $v_m$ to $E_{nr}$ one has to take into
account that different disconnected regions in $E_{nr}$ can contribute
to a given interval in $v_m$.

$(iv)$ The type of DM--nucleus interaction (i.e., the particle
physics) has to be specified before applying our bounds. Apart from
the $1/v^2$ dependence of the differential cross section
$d\sigma/dE_{nr}$, the $E_{nr}$ dependence of the interaction is
encoded in the form factor $F(E_{nr})$. It can describe conventional
spin-independent or spin-dependent interactions, but also a possible
non-trivial $q^2$-dependence of the DM--nucleus interaction, all of which can be
absorbed into $F(E_{nr})$.

\bigskip

To conclude, we have presented a consistency check for the amplitude of a DM
induced annual modulation compared to the unmodulated event rate in a given
DM direct detection experiment. Our bounds rely only on very mild and
realistic assumptions about the DM halo. We believe that the proposed method is
a useful check which any annual modulation signal should pass. A violation
of our bounds suggests a non-DM origin of the annual modulation, or requires
rather exotic properties of the DM distribution, for example very sharp spikes
and edges. Such extreme features of the halo should have other observable
consequences, such as surprising spectral features or strong energy
dependence of the modulation phase. Such features could be used as
additional diagnostics, beyond the signatures explored here, which are
restricted to the energy integrated modulation amplitude, irrespective of
the phase.

\acknowledgments

We thank Felix Kahlh\"ofer for useful comments on the manuscript.
T.S.\ and J.Z.\ are grateful for support from CERN for attending the 
CERN-TH institute DMUH'11, 18--29 July 2011, where this work was initiated.
The work of T.S.\ was partly supported by the Transregio
Sonderforschungsbereich TR27 ``Neutrinos and Beyond'' der Deutschen
Forschungsgemeinschaft. J.H.-G.\ is supported by the MICINN under the FPU
program and would like to thank the division of M.~Lindner at MPIK for very
kind hospitality.

\appendix

\section{Maxwellian halo}
\label{sec:maxw}

As a cross check we apply our formalism to the commonly
used Maxwellian velocity distribution, with a cut off
at an escape velocity $v_{\rm esc}$.  In the galactic rest frame it is given by
\beq\label{eq:fmaxw}
f_\mathrm{gal}(\vect{v}) \propto [\exp(-\vect{v}^2/\bar{v}^2) - \exp(v_{\rm
esc}^2/\bar{v}^2)]\Theta(v_{\rm esc}-v) \,, 
\eeq
where we use $\bar v = 220\,\rm km\, s^{-1}$ and $v_{\rm esc} =
550\,\rm km\, s^{-1}$. This distribution is boosted to the sun's and
earth's rest frames as described in section~\ref{sec:notation}. For
the above velocity distribution the halo integral $\eta(v_m,t)$ is
known analytically.  We define the ``exact'' modulated and unmodulated
halo integrals as
\beq\label{eq:maxwell}
\begin{split}
\bar\eta^{\rm exact}(v_m) &\equiv \frac{1}{2}
\left[\eta(v_m, t_*) + \eta(v_m, t_* + 0.5)\right]
\,,\\
{A'_\eta}^{\rm exact}(v_m) &\equiv \frac{1}{2} 
\left[\eta(v_m, t_*) - \eta(v_m, t_* + 0.5)\right]
\,,
\end{split}
\eeq
where for the (exact) analytic expressions $\eta(v_m,t)$ we are are using the results given in the appendix of
\cite{McCabe:2010zh}. Above, $t_*$ is June 2nd and $A_\eta^{\rm exact}(v_m) =
|{A'_\eta}^{\rm exact}(v_m)|$. 

\begin{figure}
\includegraphics[width=0.65\textwidth]{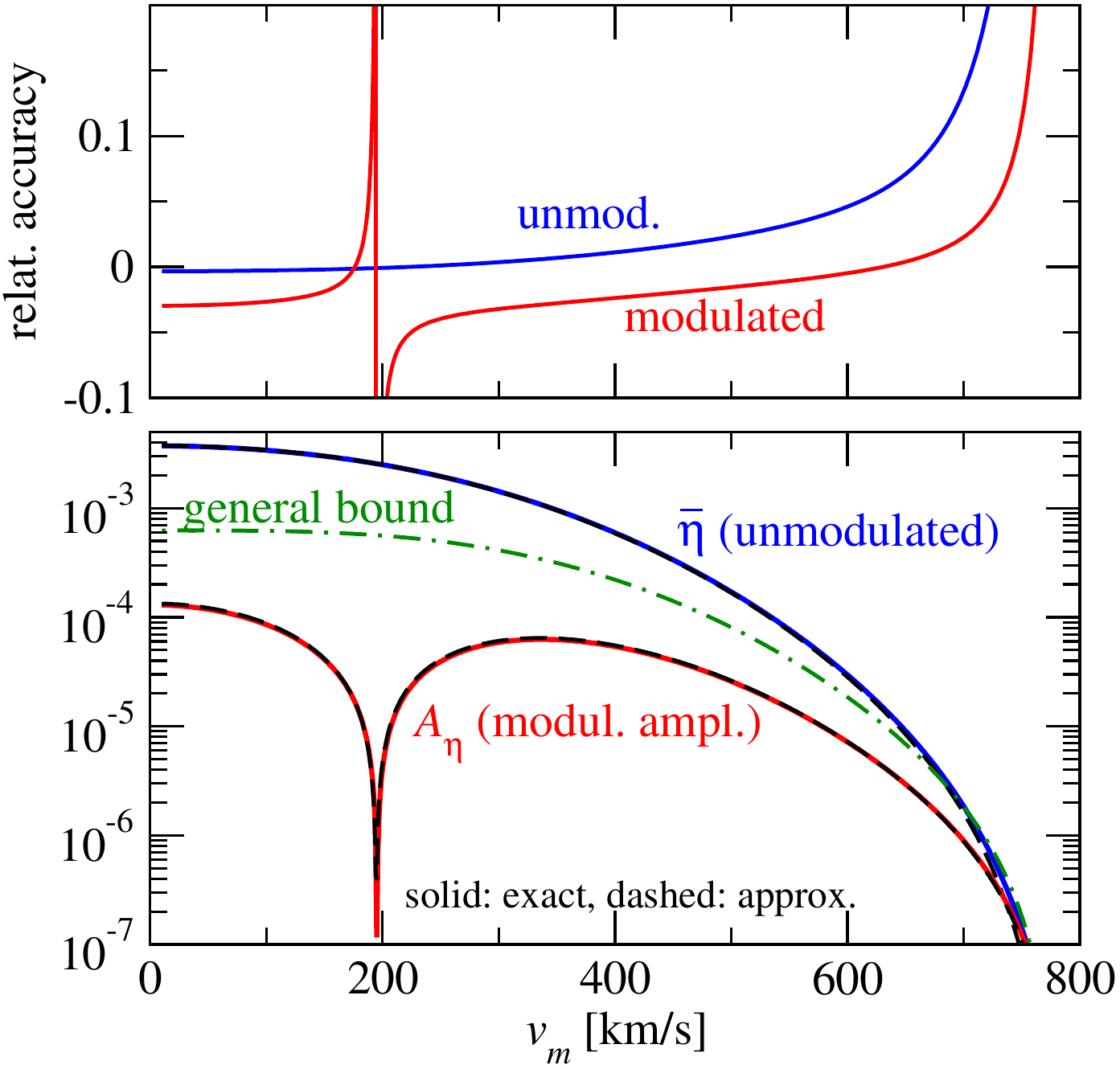}
\mycaption{\label{fig:maxwell} Comparison for the Maxwellian halo of the exact (solid line) and $v_e$ expanded (dashed) modulated and
  unmodulated halo integrals $A_\eta$ and $\bar \eta$, respectively. 
  The modulation flips the phase by half a year below the
  zero of the amplitude. In the lower panel the general bound
  \eqref{eq:bound_gen} is also shown (dashed-dotted curve). The upper
  panel shows the relative accuracy defined as (exact --
  approx)/exact.}
\end{figure}

We first check the accuracy of the $v_e$ expansion. Expanding to zeroth
order in $v_e$ gives the approximate unmodulated rate \eqref{eq:unmod},
while the modulation amplitude is given by expanding to linear order in
$v_e$, \eqref{eq:M1:special}, with $\sin\alpha_{\rm halo} = 0.5$. The
comparison with the ``exact" expressions is shown in fig.~\ref{fig:maxwell}.
Both for the modulation amplitude and the unmodulated rate we find excellent
agreement, with the differences hardly visible on logarithmic scale. As seen
from the upper panel, for large regions of $v_m$ the agreement is within few
\%, apart from the zero of the modulation amplitude. Minor deviations appear
for large $v_m$ values, where non-linear effects due to the cut-off at the
escape velocity become important. We see that our expression for $A_\eta$
captures accurately the phase flip of the modulation. At low $v_m$ the
$G(v_m)$ term in eq.~\eqref{eq:M1:special} dominates, leading to the maximum
of the count rate in December, the modulation amplitude becomes zero when
the two terms cancel, and the maximum in June at large $v_m$ is obtained
from the $g(v_m)$ term.

\begin{figure}
\includegraphics[height=0.44\textwidth]{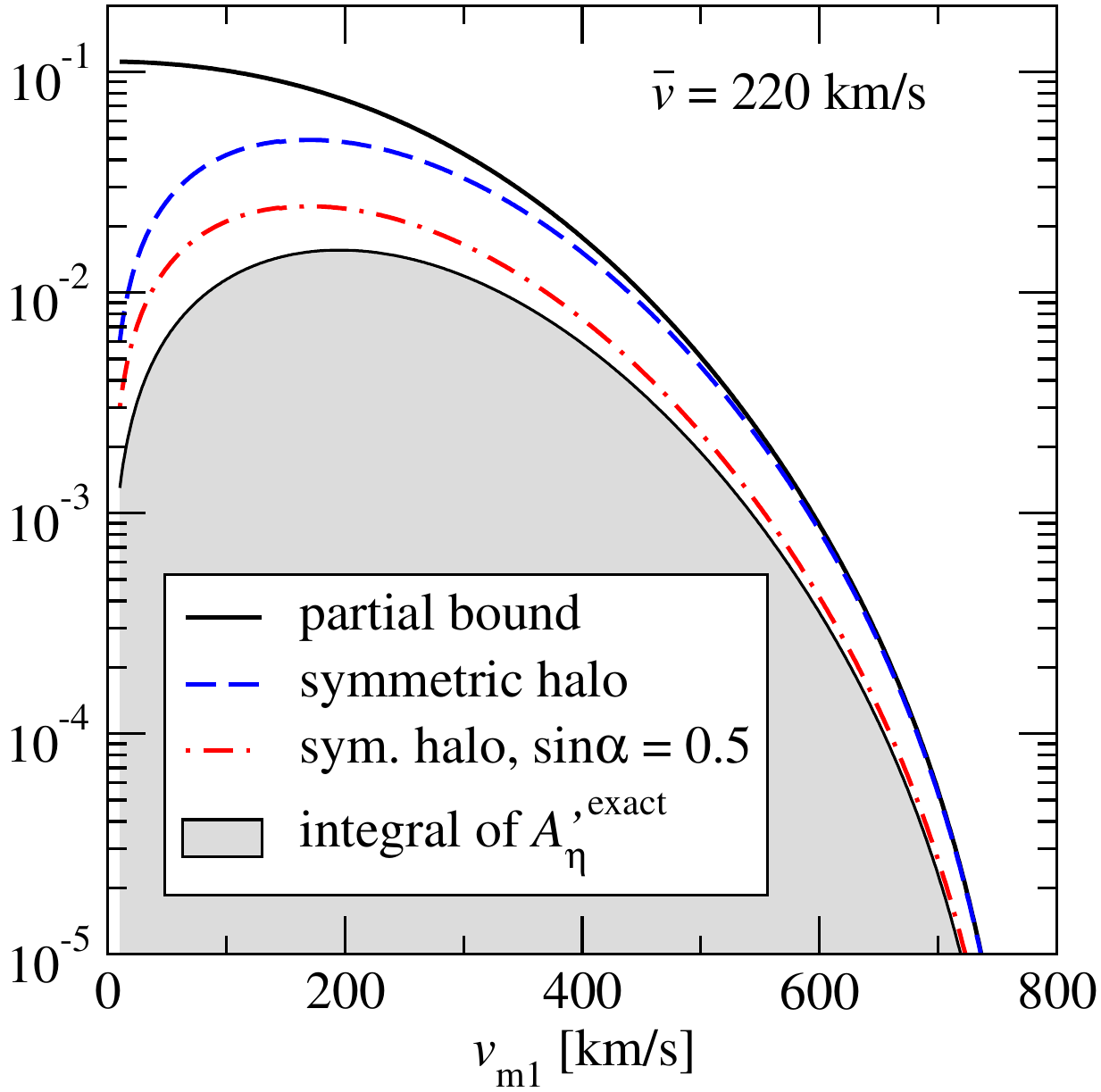}\quad
\includegraphics[height=0.44\textwidth]{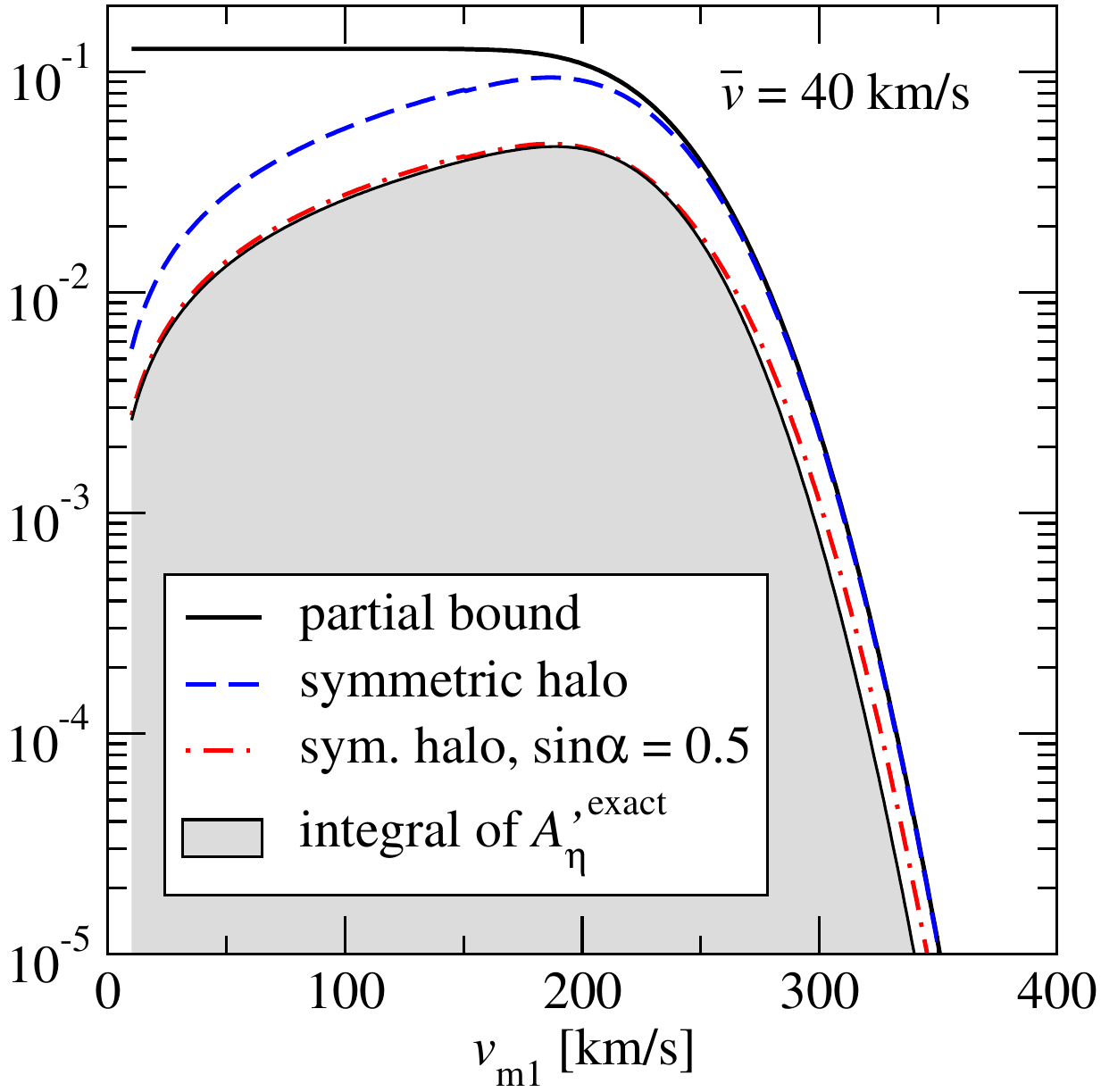}
\mycaption{\label{fig:maxwell2} The bounds on the modulation amplitude for a Maxwellian
  velocity distribution and velocity dispersions $\overline v = 220$~km/s (40~km/s) in the left (right) panels. The shaded region shows the integral
  of $\int_{v_{m1}}^\infty {A'_\eta}^{\rm exact}$. 
  We show the bound for a ``symmetric halo'' from
  eq.~\eqref{eq:bound_spec}, and the bound from
  eq.~\eqref{eq:bound_spec_alpha} which assumes that the DM
  wind is parallel to the motion of the sun
  (``sym.\ halo, $\sin\alpha = 0.5$'').  The solid curve (``partial
  bound'') shows the bound eq.~\eqref{eq:bound_spec}, but without
  the second term in the square bracket.}
\end{figure}

We compare next the bounds with the actual modulation amplitude.  The
dash-dotted (green) curve in fig.~\ref{fig:maxwell} shows the general bound,
eq.~\eqref{eq:bound_gen}, which follows from Assumption~1. Clearly, the
Maxwellian halo is far from saturating this bound. Much tighter constraints
are obtained from Assumptions~2 and 2a. Fig.~\ref{fig:maxwell2} shows the
bounds \eqref{eq:bound_spec} and \eqref{eq:bound_spec_alpha} on integrals of
modulation amplitudes. The shaded region shows the l.h.s.\ of the bounds,
$\int_{v_{m1}}dv A_\eta'$, where the upper boundary of the integration is
chosen so high that the DM signal vanishes. Note that $A_\eta'$ flips the
sign at $v_m\sim 200$ km/s, which explains the maximum of the integral
around 200~km/s.  The curve labeled ``partial bound'' corresponds to the
bound following from Assumption 2, eq.~\eqref{eq:bound_spec}, but without
the second term in the square bracket. We find that it is only slightly more
stringent than the general bound that follows from Assumption 1,
\eqref{eq:bound_gen}, especially for large $v_m$. Although it does not
capture the behaviour at low $v_m$ it is rather similar to the full bound,
eq.~\eqref{eq:bound_spec}, (dashed curve) for $v_m \gtrsim 400$~km/s. The
bound from Assumption 2a, eq.~\eqref{eq:bound_spec_alpha} (dash-dotted
curve), which assumes $\sin\alpha_{\rm halo} = 0.5$, is roughly a factor 2
larger than the prediction for small $v_{m1}$ and approaches the true
modulation around $v_{m1} \gtrsim 400$~km/s. In the right panel we show that
for a Maxwellian distribution with very small dispersion ($\overline v =
40$~km/s) the bound for Assumption~2a is nearly saturated. As mentioned in
sec.~\ref{sec:general}, the inequalities used to derive the bound
\eqref{eq:bound_spec_alpha} become equalities if the velocity distribution
$f(\vect{v}) \propto \delta(\vartheta)$, which is approximated by the
Maxwellian with small $\overline v$.

\section{Translating bounds in $v_m$ space to observable quantities}
\label{sec:translation}

In sections~\ref{sec:general} and \ref{sec:spec} we have derived
bounds involving the halo integral $\overline \eta(v_m)$ and the modulation
amplitude $A_\eta(v_m)$. Here we give details on how to translate these bounds
into bounds that involve observable rates as functions of nuclear recoil energy $E_{nr}$. 
The nuclear energy is measured in electron-equivalents, $E_{ee}$, which is 
related to $E_{nr}$ through a quenching factor $Q$. We also define a differential quenching factor $q$, 
\beq
 q \equiv \frac{d E_{ee}}{d E_{nr}} = Q + E_{nr}\frac{dQ}{d E_{nr}}\,,  \quad{\rm where}\quad Q \equiv \frac{E_{ee}}{E_{nr}}\,,
 \eeq 
which then gives the translation between $v_m$ and $E_{ee}$,
\beq
v_m = \sqrt{\frac{m_A E_{ee}}{2 \mu_{\chi A}^2 Q}}, \qquad\quad
\frac{1}{\xi(E_{ee})}\equiv \frac{dv_m}{dE_{ee}} = 
\frac{dv_m}{dE_{nr}} \frac{dE_{nr}}{dE_{ee}} = 
\frac{1}{2\mu_{\chi A}} \sqrt{\frac{m_A Q}{2 E_{ee}}} \,\frac{1}{q}\,. \label{eq:vmEee}
\eeq
The quenching factor $Q$ is in general energy dependent. If it is energy independent,
then $q=Q$, while if data are reported directly in
$E_{nr}$ then furthermore $q=Q=1$. Note that $q, Q, v_m, \xi$, are all
positive. 

The modulation amplitude and rate in units of
counts/day/kg/keV$_{ee}$ are given as
\beq
A_R(E_{ee}) = \kappa \, A_\eta(v_m) \,,\quad
\overline R(E_{ee}) = \kappa\, \overline\eta(v_m) 
\quad\text{with}\quad
\kappa(E_{ee}) \equiv \frac{C F^2(E_{ee}/Q)}{ q(E_{ee})} \,,
\eeq
with the constant $C$ defined in~\eqref{eq:Rgamma}.  To keep the
notation simple we use the same symbols for rate and modulation
whether they depend on recoil energy in keV$_{ee}$ or keV$_{nr}$. The
units are denoted in the argument ($E_{ee}$ versus $E_{nr}$) and
should be clear from the context.

Let us consider first the general bound following from Assumption~1.
Multiplying eq.~\eqref{eq:bound_gen} by $\kappa$ and converting the
integral over $v_m$ into $E_{ee}$ by using eq.~\eqref{eq:vmEee} one finds
\beq\label{eq:bound_app}
A_{R}(E_{ee}) \leq v_e \left[
-\frac{d(\overline R \xi)}{dE_{ee}} 
+\overline R \left(\frac{d\xi}{dE_{ee}} + 
        \sqrt{\frac{2\mu_{\chi A}^2 Q}{m_A E_{ee}}} +
	\frac{\xi}{\kappa}\frac{d\kappa}{dE_{ee}} \right)
- \kappa \int_{E_{ee}} dE_{ee} \frac{\overline R }{\xi \kappa}
          \frac{2\mu_{\chi A}^2 Q}{m_A E_{ee}}
\right]
\eeq
Eq.~\eqref{eq:M2:special2} for Assumptions 2, 2a becomes similarly:
\beq\label{eq:bound_app2}
A_{R}(E_{ee}) \leq v_e \sin\alpha_{\rm halo} \left[
 - \frac{d(\overline R \xi)}{dE_{ee}} 
+ \overline R \left(\frac{d\xi}{dE_{ee}} +
	\frac{\xi}{\kappa}\frac{d\kappa}{dE_{ee}} \right)\right] \,.
\eeq

Note that the constant $C$ containing the cross section and local DM density
drops out, as expected, since it is a common pre-factor for rate as well as
modulation. In practise $A_R$ and $\overline R$ are given in bins in
$E_{ee}$. Let us define the bin average of a quantity $X(E_{ee})$ as 
\beq
\ev{X}_i \equiv \frac{1}{\Delta E_i} \int_{E_{i1}}^{E_{i2}} dE_{ee} X(E_{ee}) \,,
\eeq
where $E_{i1}$ and $E_{i2}$ are the boundaries of bin $i$ and $\Delta E_i =
E_{i2} - E_{i1}$. The observed modulation and rate in bin $i$ are then $A_i = \ev{A_R}_i$
and $R_i = \ev{\overline R}_i$, respectively. Now we have to perform the bin average of
eq.~\eqref{eq:bound_app}. The first term in the square bracket gives
\beq
 \frac{\overline R(E_{i1}) \xi(E_{i1}) - \overline R(E_{i2}) \xi(E_{i2})}  {\Delta E_i}
\approx
\frac{R_i \ev{\xi}_i - R_{i+1} \ev{\xi}_{i+1}}{\Delta E_i} \,.
\eeq
For the other two terms in the square bracket we approximate $\ev{XY}
\approx \ev{X}\ev{Y}$ and the integral becomes a sum over bins, up to the
last reported bin $i=N$. Finally we obtain
\begin{align}
A_i &\leq v_e \left[
R_i (\alpha_i + \beta_i) -  R_{i+1} \alpha'_{i+1}
 - \ev{\kappa}_i \sum_{j=i}^N R_j \gamma_j \right] \,, 
\end{align}
with
\beq\label{eq:def_consts1}
\alpha_i \equiv \frac{\ev{\xi}_i}{\Delta E_i} \,,\,
\alpha'_i \equiv \frac{\ev{\xi}_i}{\Delta E_{i-1}} \,,\,
\beta_i\equiv
        \ev{\frac{d\xi}{dE_{ee}} + 
        \sqrt{\frac{2\mu_{\chi A}^2 Q}{m_A E_{ee}}} +
	\frac{\xi}{\kappa}\frac{d\kappa}{dE_{ee}}}_i  \,,\,
\gamma_i  \equiv \Delta E_i
         \ev{\frac{2\mu_{\chi A}^2 Q}{m_A E_{ee} \xi \kappa}}_i \,.
\eeq
For eq.~\eqref{eq:bound_app2} we get
\begin{align}
A_i &\leq v_e \sin\alpha_{\rm halo}\left[
R_i (\alpha_i + \beta'_i) -  R_{i+1} \alpha'_{i+1}
\right]
\quad\text{with}\quad
\beta_i'\equiv
        \ev{\frac{d\xi}{dE_{ee}} + 
	\frac{\xi}{\kappa}\frac{d\kappa}{dE_{ee}}}_i \,.
\end{align}
Terms with indices larger than $N$ are dropped. We see from the definitions
that $\alpha_i, \alpha'_i, \gamma_i$, and $\ev{\kappa}_i$ are positive.

A similar calculation for the bounds
\eqref{eq:bound_spec} and \eqref{eq:bound_spec_alpha}, following
from the Assumptions~2 and 2a, leads to
\beq
\sum_{j=i}^N A_j x_j \leq v_e \sin\alpha 
\left[
R_i y_i - \ev{v_m}_i \sum_{j=i+1}^N R_j \gamma_j 
\right],
\eeq
with
\beq\label{eq:def_consts2}
x_i \equiv \Delta E_i \ev{\frac{1}{\xi \kappa}}_i \,,\qquad
y_i \equiv \ev{\frac{1}{\kappa}}_i - \ev{v_m}_i\gamma_i  \,,\qquad
\ev{v_m}_i = \ev{\sqrt{\frac{m_A E_{ee}}{2 \mu_{\chi A}^2 Q}}}_i \,,
\eeq
and $\sin\alpha = 1\,(0.5)$ corresponds to Assumption~2 (2a).

Any binning procedure will lead to inaccuracies. We have estimated binning
effects by slightly changing the prescription. Eq.~\eqref{eq:bound_app}
can be equivalently written as
\beq
A_{R}(E_{ee}) \leq v_e \left[
- \xi \frac{d\overline R}{dE_{ee}} 
+\overline R \left(\sqrt{\frac{2\mu_{\chi A}^2 Q}{m_A E_{ee}}} +
	\frac{\xi}{\kappa}\frac{d\kappa}{dE_{ee}} \right)
- \kappa \int_{E_{ee}} dE_{ee} \frac{\overline R }{\xi \kappa}
          \frac{2\mu_{\chi A}^2 Q}{m_A E_{ee}}
\right] \,.
\eeq
The first term in the square bracket can be estimated as 
\beq
 - \xi \frac{d\overline{R}}{dE_{ee}}
\approx
\langle\xi\rangle_i \frac{R_i- R_{i+1}}{\Delta E_i} \,,
\eeq
and the other terms are bin-averaged similar as above. We have calculated
the CoGeNT bounds also using this binning procedure and obtain similar
results as with our default method. This confirms that inaccuracies due to
binning are acceptable for present CoGeNT data.

\section{Alternative procedure for optimizing the bound in presence of unknown
background}
\label{app:proc2}

In sec.~\ref{sec:data} we have outlined a method based on a
least-square minimization (``procedure 2'') to find ``optimal''
values for the coefficients $\omega_i$ parameterizing the background
contribution in each bin. Here we provide an alternative method for
this task. 

Let us consider the inequalities~\eqref{eq:gen_bin1} and
\eqref{eq:sym_bin1} as a system of equations for the $\omega_i$. We
observe that they have a triangular structure with respect to
$\omega_i$: the inequality for a given index $i$ depends only on
$\omega_j$ with $j \ge i$, i.e., the inequality for $i=N$ depends only
on $\omega_N$, for $i=N-1$ it depends on $\omega_{N-1}, \omega_N$, and
so on. From the fact that $\alpha_i, \ev{\kappa}_i, \gamma_i$ are
positive follows that the bound for a given $i$ is weakest if
$\omega_i$ is as large as possible and $\omega_j$ with $j > i$ are as
small as possible. We can implement this requirement in an iterative
way: replacing the inequality sign by an equality, we obtain a system
of $N$ linear equations in $\omega_i$. We can solve this system by
starting at $i=N$ and going up step by step to $i=1$, in each step
obtaining a value for the corresponding $\omega_i$, to be used in the
next step. If the solution for $\omega_i$ is less than one, it will be
the smallest allowed value, and hence the bound for $i-1$ will be
weakest. If the solution for $\omega_i$ is larger than one the bound
is violated by bin $i$ and we have to set the corresponding $\omega_i
= 1$ (the largest physically allowed value). In that way we obtain a
set of $\omega_i$ corresponding to the most conservative choice of
background contributions to the unmodulated rate.

\begin{table}
\begin{tabular}{ll|c|c|c|c|c|c}
\hline\hline
& & \multicolumn{3}{|c|}{unmod.\ rate from \cite{Aalseth:2011wp}} &
    \multicolumn{3}{|c }{corrected rate \cite{Collar-TAUP}} \\
\hline    
& & {Mean} & {68\%} & {95\%} 
  & {Mean} & {68\%} & {95\%} \\
\hline
Ass.\ 1:  & general bound  & 10 &  5 & $-$ &
                             12.5 &  5 & $-$  \\
Ass.\ 2:  & symmetric halo & 29.5 &  17  & 7 &
                             63 &  34.5  & 16 \\
Ass.\ 2a: & sym.\ halo, $\sin\alpha = 0.5$ & 
37.5 & 14  & $-$ &
94.5 & 30.5  & $-$ \\
\hline\hline
\end{tabular}
  \mycaption{\label{tab:bounds-app} Lower bounds on the DM mass in GeV, from
    the requirement that the CoGeNT modulation amplitude is consistent with
    the upper bound from the unmodulated rate, according to the Assumptions
    1, 2, 2a on the DM distribution obtained as described in
    appendix~\ref{app:proc2}. In the left part of the table we use the
    published unmodulated event rates from \cite{Aalseth:2011wp}, whereas
    for the right part of the table we adopt the preliminary results on 
    surface events contamination at low energies from \cite{Collar-TAUP}.
    }
\end{table}

This method treats the inequalities as ``hard bounds'' and does not
take into account the fact that the amplitudes are subject to
experimental uncertainties. A conservative way to include
uncertainties is to apply this procedure as described above but using
instead of the l.h.s.\ of \eqref{eq:gen_bin1} and \eqref{eq:sym_bin1}
its central value minus $n$ standard deviations. In this way one aims
to satisfy the inequalities within the $n\sigma$ interval for each
bin. We show the results of such an analysis for CoGeNT data in
tab.~\ref{tab:bounds-app}. Typically one finds qualitatively similar
but slightly stronger bounds than for procedure~1, although if the
method is considered at higher CL bounds may become weaker or even
disappear, as visible in the table for the columns at 95\%~CL.
Note, however, that subtracting $n\sigma$ from the mean value for
all bins leads to very conservative limits.

\bibliographystyle{my-h-physrev.bst}
\bibliography{./refs}

\begin{thebibliography}{10}

\bibitem{Drukier:1986tm}
A.~K. Drukier, K.~Freese, and D.~N. Spergel,
\newblock {\em {Detecting Cold Dark Matter Candidates}},
\newblock Phys. Rev. {\bf D33}, 3495 (1986).

\bibitem{Freese:1987wu}
K.~Freese, J.~A. Frieman, and A.~Gould,
\newblock {\em {Signal Modulation in Cold Dark Matter Detection}},
\newblock Phys. Rev. {\bf D37}, 3388 (1988).

\bibitem{Bernabei:2008yi}
DAMA, R.~Bernabei {\em et~al.},
\newblock {\em {First results from DAMA/LIBRA and the combined results with
  DAMA/NaI}},
\newblock Eur. Phys. J. {\bf C56}, 333 (2008), 0804.2741.

\bibitem{Aalseth:2011wp}
C.~E. Aalseth {\em et~al.},
\newblock {\em {Search for an Annual Modulation in a P-type Point Contact
  Germanium Dark Matter Detector}},
\newblock (2011), 1106.0650.

\bibitem{Kuhlen:2009vh}
M.~Kuhlen {\em et~al.},
\newblock {\em {Dark Matter Direct Detection with Non-Maxwellian Velocity
  Structure}},
\newblock JCAP {\bf 1002}, 030 (2010), 0912.2358.

\bibitem{Fornengo:2003fm}
N.~Fornengo and S.~Scopel,
\newblock {\em {Temporal distortion of annual modulation at low recoil
  energies}},
\newblock Phys. Lett. {\bf B576}, 189 (2003), hep-ph/0301132.

\bibitem{Green:2003yh}
A.~M. Green,
\newblock {\em {Effect of realistic astrophysical inputs on the phase and shape
  of the WIMP annual modulation signal}},
\newblock Phys. Rev. {\bf D68}, 023004 (2003), astro-ph/0304446.

\bibitem{Fairbairn:2008gz}
M.~Fairbairn and T.~Schwetz,
\newblock {\em {Spin-independent elastic WIMP scattering and the DAMA annual
  modulation signal}},
\newblock JCAP {\bf 0901}, 037 (2009), 0808.0704.

\bibitem{Gelmini:2000dm}
G.~Gelmini and P.~Gondolo,
\newblock {\em {WIMP annual modulation with opposite phase in Late-Infall halo
  models}},
\newblock Phys.Rev. {\bf D64}, 023504 (2001), hep-ph/0012315.

\bibitem{Savage:2006qr}
C.~Savage, K.~Freese, and P.~Gondolo,
\newblock {\em {Annual Modulation of Dark Matter in the Presence of Streams}},
\newblock Phys. Rev. {\bf D74}, 043531 (2006), astro-ph/0607121.

\bibitem{Chang:2008xa}
S.~Chang, A.~Pierce, and N.~Weiner,
\newblock {\em {Using the Energy Spectrum at DAMA/LIBRA to Probe Light Dark
  Matter}},
\newblock Phys.Rev. {\bf D79}, 115011 (2009), 0808.0196.

\bibitem{Natarajan:2011gz}
A.~Natarajan, C.~Savage, and K.~Freese,
\newblock {\em {Probing dark matter streams with CoGeNT}},
\newblock (2011), 1109.0014.

\bibitem{Drees:2007hr}
M.~Drees and C.-L. Shan,
\newblock {\em {Reconstructing the Velocity Distribution of WIMPs from Direct
  Dark Matter Detection Data}},
\newblock JCAP {\bf 0706}, 011 (2007), astro-ph/0703651.

\bibitem{Drees:2008bv}
M.~Drees and C.-L. Shan,
\newblock {\em {Model-Independent Determination of the WIMP Mass from Direct
  Dark Matter Detection Data}},
\newblock JCAP {\bf 0806}, 012 (2008), 0803.4477.

\bibitem{Fox:2010bz}
P.~J. Fox, J.~Liu, and N.~Weiner,
\newblock {\em {Integrating Out Astrophysical Uncertainties}},
\newblock Phys.Rev. {\bf D83}, 103514 (2011), 1011.1915.

\bibitem{Fox:2010bu}
P.~J. Fox, G.~D. Kribs, and T.~M. Tait,
\newblock {\em {Interpreting Dark Matter Direct Detection Independently of the
  Local Velocity and Density Distribution}},
\newblock Phys.Rev. {\bf D83}, 034007 (2011), 1011.1910.

\bibitem{McCabe:2011sr}
C.~McCabe,
\newblock {\em {DAMA and CoGeNT without astrophysical uncertainties}},
\newblock (2011), 1107.0741.

\bibitem{Frandsen:2011gi}
M.~T. Frandsen, F.~Kahlhoefer, C.~McCabe, S.~Sarkar, and K.~Schmidt-Hoberg,
\newblock {\em {Resolving astrophysical uncertainties in dark matter direct
  detection}},
\newblock (2011), 1111.0292.

\bibitem{Read:2009iv}
J.~Read, L.~Mayer, A.~Brooks, F.~Governato, and G.~Lake,
\newblock {\em {A dark matter disc in three cosmological simulations of Milky
  Way mass galaxies}},
\newblock (2009), 0902.0009.

\bibitem{Fox:2011px}
P.~J. Fox, J.~Kopp, M.~Lisanti, and N.~Weiner,
\newblock {\em {A CoGeNT Modulation Analysis}},
\newblock (2011), 1107.0717.

\bibitem{Collar-TAUP}
J.~Collar,
\newblock {CoGeNT and COUPP},
\newblock talk at TAUP 2011, "12th International Conference on Topics in
  Astroparticle and Underground Physics", 2011.

\bibitem{Frandsen:2011ts}
M.~T. Frandsen {\em et~al.},
\newblock {\em {On the DAMA and CoGeNT Modulations}},
\newblock Phys.Rev. {\bf D84}, 041301 (2011), 1105.3734.

\bibitem{Schwetz:2011xm}
T.~Schwetz and J.~Zupan,
\newblock {\em {Dark Matter attempts for CoGeNT and DAMA}},
\newblock JCAP {\bf 1108}, 008 (2011), 1106.6241.

\bibitem{Chang:2011eb}
S.~Chang, J.~Pradler, and I.~Yavin,
\newblock {\em {Statistical Tests of Noise and Harmony in Dark Matter
  Modulation Signals}},
\newblock (2011), 1111.4222.

\bibitem{Arina:2011zh}
C.~Arina, J.~Hamann, R.~Trotta, and Y.~Y. Wong,
\newblock {\em {Evidence for dark matter modulation in CoGeNT}},
\newblock (2011), 1111.3238.

\bibitem{Barbeau:2007qi}
P.~S. Barbeau, J.~I. Collar, and O.~Tench,
\newblock {\em {Large-Mass Ultra-Low Noise Germanium Detectors: Performance and
  Applications in Neutrino and Astroparticle Physics}},
\newblock JCAP {\bf 0709}, 009 (2007), nucl-ex/0701012.

\bibitem{McCabe:2010zh}
C.~McCabe,
\newblock {\em {The Astrophysical Uncertainties Of Dark Matter Direct Detection
  Experiments}},
\newblock Phys. Rev. {\bf D82}, 023530 (2010), 1005.0579.

\end{thebibliography}

\end{document}